\documentclass[english,aps, preprint, floatfix]{revtex4-1}
\usepackage[latin9]{inputenc}
\usepackage{graphicx}

\makeatletter

\providecommand{\tabularnewline}{\\}

\@ifundefined{textcolor}{}
{%
 \definecolor{BLACK}{gray}{0}
 \definecolor{WHITE}{gray}{1}
 \definecolor{RED}{rgb}{1,0,0}
 \definecolor{GREEN}{rgb}{0,1,0}
 \definecolor{BLUE}{rgb}{0,0,1}
 \definecolor{CYAN}{cmyk}{1,0,0,0}
 \definecolor{MAGENTA}{cmyk}{0,1,0,0}
 \definecolor{YELLOW}{cmyk}{0,0,1,0}
}

\usepackage{babel}

\makeatother

\usepackage{babel}
\begin{document}

\title{Probing Charged Higgs Boson Couplings at the FCC-hh Collider}

\author{I.T. Cakir}

\email{ilkayturkcakir@aydin.edu.tr}

\affiliation{Istanbul Aydin University, Application and Research Center for Advanced
Studies, 34295 Sefakoy, Istanbul, Turkey}

\author{S. Kuday}

\email{sinankuday@aydin.edu.tr}

\affiliation{Istanbul Aydin University, Application and Research Center for Advanced
Studies, 34295 Sefakoy, Istanbul, Turkey}

\author{H. Saygin}

\email{hasansaygin@aydin.edu.tr}

\affiliation{Istanbul Aydin University, Application and Research Center for Advanced
Studies, 34295 Sefakoy, Istanbul, Turkey}

\author{A. Senol}

\email{senol_a@ibu.edu.tr}

\affiliation{Abant Izzet Baysal University, Department of Physics, 14280 Golkoy,
Bolu, Turkey}

\author{O. Cakir}

\email{orhancakir@aydin.edu.tr}

\email{ocakir@science.ankara.edu.tr}

\affiliation{Istanbul Aydin University, Application and Research Center for Advanced
Studies, 34295 Sefakoy, Istanbul, Turkey}

\affiliation{Ankara University, Department of Physics, 06100 Tandogan, Ankara,
Turkey}

\date{\today}
\begin{abstract}
Many of the new physics models predicts a light Higgs boson similar
to the Higgs boson of the Standard Model (SM) and also extra scalar
bosons. Beyond the search channels for a SM Higgs boson, the future
collider experiments will explore additional channels that are specific
to extended Higgs sectors. We study the charged Higgs boson production
within the framework of two Higgs doublet models (THDM) in the proton-proton
collisions at the FCC-hh collider. With an integrated luminosity of
500 fb$^{-1}$ at very high energy frontier, we obtain a significant
coverage of the parameter space and distinguish the charged Higgs-top-bottom
interaction within the THDM or other new physics models with charged
Higgs boson mass up to 1 TeV.
\end{abstract}

\pacs{14.80.Cp, 14.65.Ha, 12.60.-i}

\maketitle

\section{Introduction}

The Higgs sector of the Standard Model (SM) is in the minimal form,
which contains one complex isospin doublet of scalar fields ($\phi^{+},\phi^{0}$),
resulting in one physical neutral CP-even Higgs boson ($h^{0}$) which
has been discovered at the LHC by ATLAS \cite{Aad2012} and CMS \cite{Chatrchyan2012}
collaborations. However, there are many possibilities for the extension
of the Higgs sector, introducing further multiplets of scalar fields,
which might be singlets, doublets, or triplets of the symmetry groups.
The extended Higgs sectors are more relevant to the neutrino mass,
baryogenesis and dark matter. It is thus natural to consider scenarios
with additional complex scalars such as two Higgs doublet model (THDM)
\cite{Branco2012}. The scalar fields ($\phi_{1}$ and $\phi_{2}$)
couplings to the up-type, down-type and charged lepton $SU(2)_{L}$
singlet fermions can be identified for discrete types of the THDM
\cite{PDG2014,Gunion2000}. Model of type-I are the one in which all
SM fermions couple to a single scalar field ($\phi_{2}$). In type-II
model down-type quarks and charged leptons couple to one scalar field
($\phi_{1}$), while the up-type quarks and neutrinos couple to the
other scalar field ($\phi_{2}$). Furthermore, in the model of type-III
the quarks couple to one of the scalar field ($\phi_{1}$), while
leptons couple to the other ($\phi_{2}$). In the model of type-IV,
the couplings of scalar ($\phi_{2}$) to up-type quarks and charged
leptons, and the couplings of scalar ($\phi_{1}$) to down-type quarks
and neutrinos are present. Here, we consider two types of THDM-I and
THDM-II for type-I and type-II of two Higgs doublet model, respectively.

The two Higgs doublets carry opposite hypercharges, the ($\phi_{1}$
and $\phi_{2}$) scalar potential will contain the mixing parameters
related to the mass. In this case, the Higgs doublets will have different
vacuum expectations values ($v_{1}$ and $v_{2}$). The massive SM
gauge bosons acquire their masses from the expressions with the vacuum
expectation value $v=(v_{1}^{2}+v_{2}^{2})^{1/2}$. After the spontaneous
symmetry breaking there appears five physical scalar particles: two
neutral CP-even bosons $h^{0}$ and $H^{0}$, one neutral CP-odd boson
$A^{0}$, and two charged bosons $H^{\pm}$. Phenomenologically, the
two Higgs doublet model includes the free parameters: mixing angle,
the ratio of the vacuum expectation values ($\tan\beta=v_{2}/v_{1}$),
the masses of Higgs bosons. In the extended models of multiple neutral
scalar bosons, the mixing between them would make it difficult to
identify their properties. Therefore, it is important to study the
charged scalar bosons, which could provide unique signatures to distinguish
the models with extended Higgs sector.

Constraints on the charged Higgs bosons in the THDM are given from
both low energy flavour experiments and high energy collider experiments.
The direct lower bounds on the charged Higgs boson mass $m_{H^{\pm}}$
come from LEP experiments. They were sensitive to the masses of charged
Higgs boson up to about 90 GeV, in two decay channels of $H^{+}\to\tau^{+}\nu$
and $H^{+}\to c\bar{s}$, and the exclusion limit on the mass independent
of the admixture of these branching fractions was $78.9$ GeV \cite{PDG2014}.
Assuming the THDM-II model the lower bounds are $87.8$ GeV for large
$\tan\beta$ values using the decay mode $H^{+}\to\tau^{+}\nu$ and
$80.4$ GeV for relatively low $\tan\beta$ values. Using the bounds
$m_{A^{0}}>92$ GeV and the characteristic relation $m_{H^{\pm}}=(m_{A^{0}}^{2}+m_{W^{\pm}})^{1/2}$
(as in the MSSM), one obtains the bound $m_{H^{\pm}}>122$ GeV. In
the model THDM-II the mass of charged Higgs boson is constrained by
the precision measurements of the radiative decay of $B\to X_{s}\gamma$
by the low energy experiments. The lower bound on the mass $m_{H^{\pm}}>295$
GeV are given for the model THDM-II. The decay $B\to\tau\nu$ can
also be used to constrain the charged Higgs parameters, being sensitive
to $(\tan\beta/m_{H^{\pm}})^{2}$, which yields a lower bound $m_{H^{\pm}}>300$
GeV for $\tan\beta>40$. A study of data analysis based on $b\to s\gamma$
\cite{Hermann12} have excluded a mass range up to $380$ GeV in a
variety of interesting processes and BSM scenarios. However, these
bounds can be relaxed if the MSSM or other new physics models contributes
through the loop diagrams. 

Beyond the search channels for a Standard Model Higgs boson, the LHC
experiments are exploring additional channels that are specific to
extended Higgs bosons. ATLAS \cite{ATL2013} and CMS \cite{CMS2013}
collaborations have already performed a number of extended Higgs searches
which exclude $\tan\beta>50$ in the range of heavy charged Higgs,
$m_{H^{\pm}}>200$ GeV. Recently, the CMS collaboration \cite{CMS2014}
have put $95\%$ C.L. exclusion limit on the mass of the charged Higgs
boson in 180-600 GeV mass range. This search is performed at a center
of mass energy of $8$ TeV with $19.7$ fb$^{-1}$ of data from $pp\to\bar{t}(b)H^{+}$
and $pp\to t(\bar{b})H^{-}$ production processes with $H^{+}\to\tau^{+}\nu_{\tau}$
decay mode. The ATLAS collaboration \cite{ATLAS2015} have searched
for charged Higgs bosons decaying through $H^{\pm}\to\tau^{\pm}\nu_{\tau}$
process using the proton-proton collision data at $\sqrt{s}=8$ TeV
with $19.5$ fb$^{-1}$, the results exclude a large range of $\tan\beta$
values for charged Higgs boson masses in the range $80-160$ GeV,
and exclude parameter space with high $\tan\beta$ for the range of
mass $m_{H^{\pm}}=200-250$ GeV.

The searches of the heavy Higgs bosons of the THDM have special challenge
at present high energy colliders. One of the future international
projects currently under consideration is the Future Circular Collider
(FCC) \cite{FCC} which has the potential to search for a wide parameter
range of new physics. The FCC-hh collider is to provide proton-proton
collisions at nearly an order of magnitude higher energy than the
LHC, the proposed centre-of-mass energy is $100$ TeV and the peak
luminosity is $5\times10^{34}$ cm$^{-2}$s$^{-1}$ \cite{Ball2014}.

In this work, we study the processes $pp\to t\bar{t}b(\bar{b})+X$
for the signal and background at the FCC-hh collider. Our analysis
is focused on the production of a pair of top quarks and associated
bottom quark for the charged Higgs boson search within the THDM-I
and THDM-II in the $pp$ collisions at very high energy frontier.
We have obtained the significant coverage of the parameter space at
large integrated luminosity projections for the FCC-hh collider. We
define the relevant expressions for the $H^{-}q\bar{q}$ and $H^{-}l^{+}\bar{\nu}$
couplings as well as scalar-vector-scalar ($H^{-}W^{+}h^{0}$, $H^{-}W^{+}H^{0}$,
$H^{-}W^{+}A^{0}$) couplings in section II. We present the calculation
for the decay widths and branchings ratios of the charged Higgs boson
in section III. In section IV, we plot the production cross section
according to the mass of charged Higgs boson for two types of THDM
at the center of mass energy $\sqrt{s}=100$ TeV of the $pp$ collisions.
Finally, the results from the analysis of the signal and background
are given in section V.

\section{Charged Higgs Boson Couplings}

The magnitude of the couplings for $H^{-}f_{i}\bar{f}_{j}$ interactions
are given by

\begin{equation}
g_{H^{-}q_{i}\bar{q}_{j}}\equiv g|V_{q_{i}q_{j}}|[m_{q_{i}}\cot\beta(1+\gamma_{5})+T_{x}m_{q_{j}}\tan\beta(1-\gamma_{5})]/(2\sqrt{2}m_{W})\label{eq:1}
\end{equation}

\begin{equation}
g_{H^{-}\nu_{i}l_{j}^{+}}\equiv g|U_{\nu{}_{i}l_{j}}|[T_{x}m_{l_{j}}\tan\beta(1-\gamma_{5})]/(2\sqrt{2}m_{W})\label{eq:2}
\end{equation}
where $V_{q_{i}q_{j}}$ and $U_{\nu_{i}l_{j}}$ are the CKM matrix
elements in quark sector and PMNS matrix elements in lepton sector,
respectively. The $g$ is the weak coupling constant, the $\tan\beta$
is the ratio of vacuum expectation vales of the Higgs doublets. The
$m_{q}$ and $m_{l}$ are the corresponding quark and lepton masses,
respectively. We use the parameter $T_{x}$ to identify the type of
THDM such that $T_{I}=-\cot\beta/\tan\beta$ denotes THDM-I and $T_{II}=1$
denotes THDM-II.

The couplings for $H^{-}W^{+}h^{0}$, $H^{-}W^{+}H^{0}$ and $H^{-}W^{+}A^{0}$
interactions can be written by

\begin{equation}
g_{H^{-}W^{+}h^{0}}\equiv g\cos(\beta-\alpha)(p_{H^{-}}+p_{h^{0}})/2\label{eq:3}
\end{equation}

\begin{equation}
g_{H^{-}W^{+}H^{0}}\equiv g\sin(\beta-\alpha)(p_{H^{-}}+p_{H^{0}})/2\label{eq:4}
\end{equation}

\begin{equation}
g_{H^{-}W^{+}A^{0}}\equiv ig(p_{H^{-}}+p_{A^{0}})/2\label{eq:5}
\end{equation}
where the $p_{h^{0}}$, $p_{H^{0}}$, $p_{A^{0}}$ and $p_{H^{-}}$
are the four-momenta for neutral Higgs bosons and charged Higgs boson,
respectively. We use the expression for the angle factor of $\cos^{2}(\beta-\alpha)=[(m_{h^{0}}^{2}/m_{Z}^{2})(m_{h^{0}}^{2}/m_{Z}^{2}-1)]/[(m_{H^{0}}^{2}/m_{Z}^{2}-m_{h^{0}}^{2}/m_{Z}^{2})(m_{H^{0}}^{2}/m_{Z}^{2}+m_{h^{0}}^{2}/m_{Z}^{2}-1)]$.
By implementing the relevant interaction vertices into the CalcHEP
\cite{Belyaev2013} package, we calculate the decay width and branching
ratios as well as the process cross section for different parametrizations
(PI for $m_{A^{0}}=100$ GeV and $m_{H^{0}}=m_{H^{-}}$ , PII for
$m_{A^{0}}=m_{H^{0}}=m_{H^{-}}$) within the THDM-I and THDM-II.

\section{Decay Width and Branching Ratios}

The partial decay widths of charged Higgs boson into fermionic channels
can be calculated as

\begin{eqnarray}
\Gamma(H^{-} & \to & q_{i}\bar{q}_{j})=\frac{3g^{2}\lambda^{1/2}(m_{H^{-}}^{2},m_{q_{i}}^{2},m_{q_{j}}^{2})}{32\pi m_{W}^{2}m_{H^{-}}^{3}}|V_{q_{i}q_{j}}|^{2}\nonumber \\
 &  & \times\left[(m_{H^{-}}^{2}-m_{q_{j}}^{2}-m_{q_{i}}^{2})(T_{x}^{2}m_{q_{j}}^{2}\tan^{2}\beta+m_{q_{i}}^{2}\cot^{2}\beta)-4T_{x}m_{q_{j}}^{2}m_{q_{i}}^{2}\right]\label{eq:6}
\end{eqnarray}

\begin{equation}
\Gamma(H^{-}\to\bar{\nu}_{i}l_{j}^{-})=\frac{g^{2}\lambda^{1/2}(m_{H^{-}}^{2},m_{l_{j}}^{2},0)}{32\pi m_{W}^{2}m_{H^{-}}^{3}}|U_{\nu_{i}l_{j}}|^{2}\left[T_{x}^{2}(m_{H^{-}}^{2}-m_{l_{j}}^{2})m_{l_{j}}^{2}\tan^{2}\beta\right]\label{eq:7}
\end{equation}
where $\lambda^{1/2}$ is a kinematic factor of mass squared dimension
and it can be defined as $\lambda(m_{1}^{2},m_{2}^{2},m_{3}^{2})=[(m_{1}^{2}+m_{2}^{2}-m_{3}^{2})^{2}-4m_{1}^{2}m_{2}^{2}]$.
For the channels $H^{-}\to W^{-}h^{0}$, $H^{-}\to W^{-}H^{0}$ and
$H^{-}\to W^{-}A^{0}$ we calculate the decay widths 
\begin{eqnarray}
\Gamma(H^{-} & \to & W^{-}h^{0})=\frac{g^{2}\lambda^{1/2}(m_{H^{-}}^{2},m_{W}^{2},m_{h^{0}}^{2})\cos^{2}(\beta-\alpha)}{64\pi m_{H^{-}}^{3}}\nonumber \\
 &  & \times\left[m_{W}^{2}-2(m_{H^{-}}^{2}+m_{h^{0}}^{2})+\frac{(m_{H^{-}}^{2}-m_{h^{0}}^{2})^{2}}{m_{W}^{2}}\right]\label{eq:8}
\end{eqnarray}

\begin{eqnarray}
\Gamma(H^{-} & \to & W^{-}H^{0})=\frac{g^{2}\lambda^{1/2}(m_{H^{-}}^{2},m_{W}^{2},m_{H^{0}}^{2})\sin^{2}(\beta-\alpha)}{64\pi m_{H^{-}}^{3}}\nonumber \\
 &  & \times\left[m_{W}^{2}-2(m_{H^{-}}^{2}+m_{H^{0}}^{2})+\frac{(m_{H^{-}}^{2}-m_{H^{0}}^{2})^{2}}{m_{W}^{2}}\right]\label{eq:9}
\end{eqnarray}

\begin{eqnarray}
\Gamma(H^{-} & \to & W^{-}A^{0})=\frac{g^{2}\lambda^{1/2}(m_{H^{-}}^{2},m_{W}^{2},m_{A^{0}}^{2})}{64\pi m_{H^{-}}^{3}}\nonumber \\
 &  & \times\left[m_{W}^{2}-2(m_{H^{-}}^{2}+m_{A^{0}}^{2})+\frac{(m_{H^{-}}^{2}-m_{A^{0}}^{2})^{2}}{m_{W}^{2}}\right]\label{eq:10}
\end{eqnarray}

\begin{figure}
\includegraphics[scale=0.8]{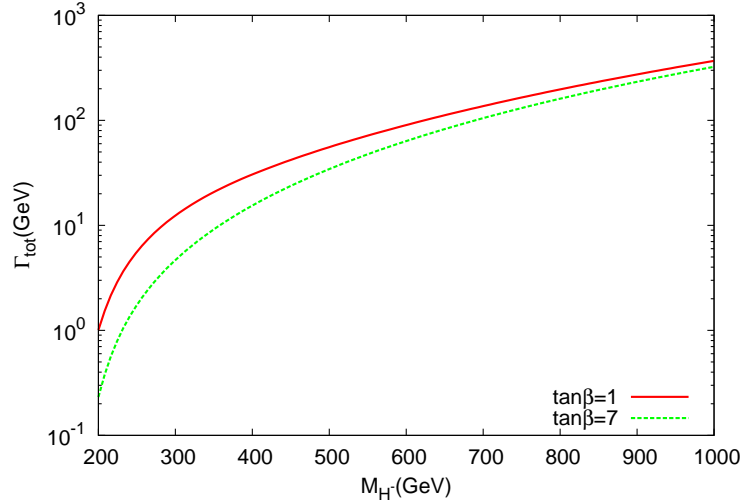}

\protect\caption{Decay width of $H^{\pm}$ boson depending on its mass for different
values of $\tan\beta$ and parameter set PI of THDM-I. \label{fig:fig1}}
\end{figure}

\begin{figure}
\includegraphics[scale=0.8]{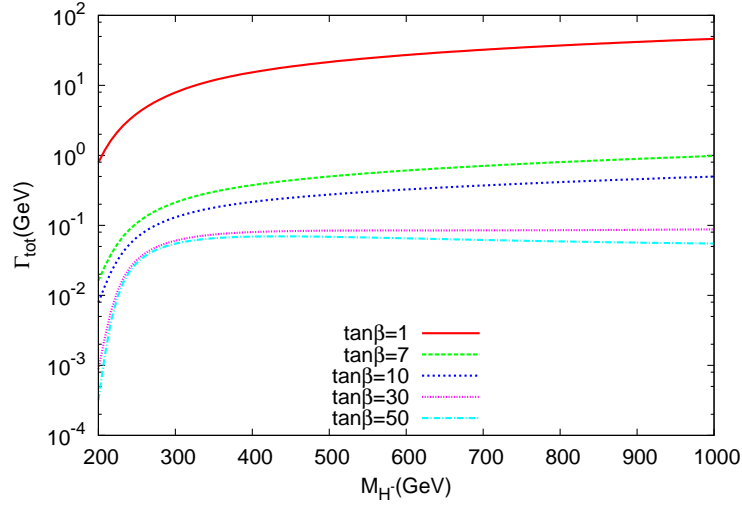}

\protect\caption{Decay width of $H^{\pm}$ boson depending on its mass for parameter
set PII of model THDM-I. \label{fig:fig2}}
\end{figure}

\begin{figure}
\includegraphics[scale=0.8]{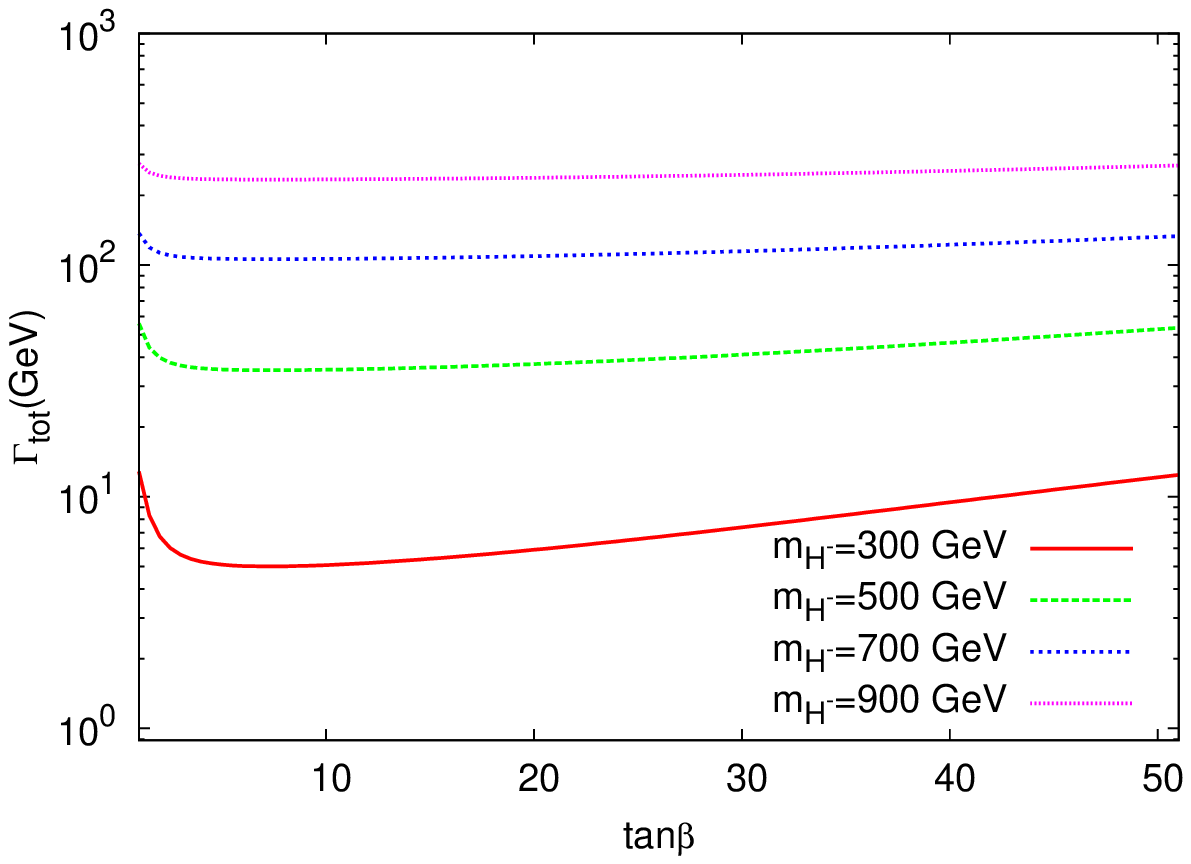}

\protect\caption{Decay width of $H^{\pm}$ boson depending on $\tan\beta$ for different
mass of charged Higgs boson for parameter set PI of model THDM-II.
\label{fig:fig3}}
\end{figure}

\begin{figure}
\includegraphics[scale=0.8]{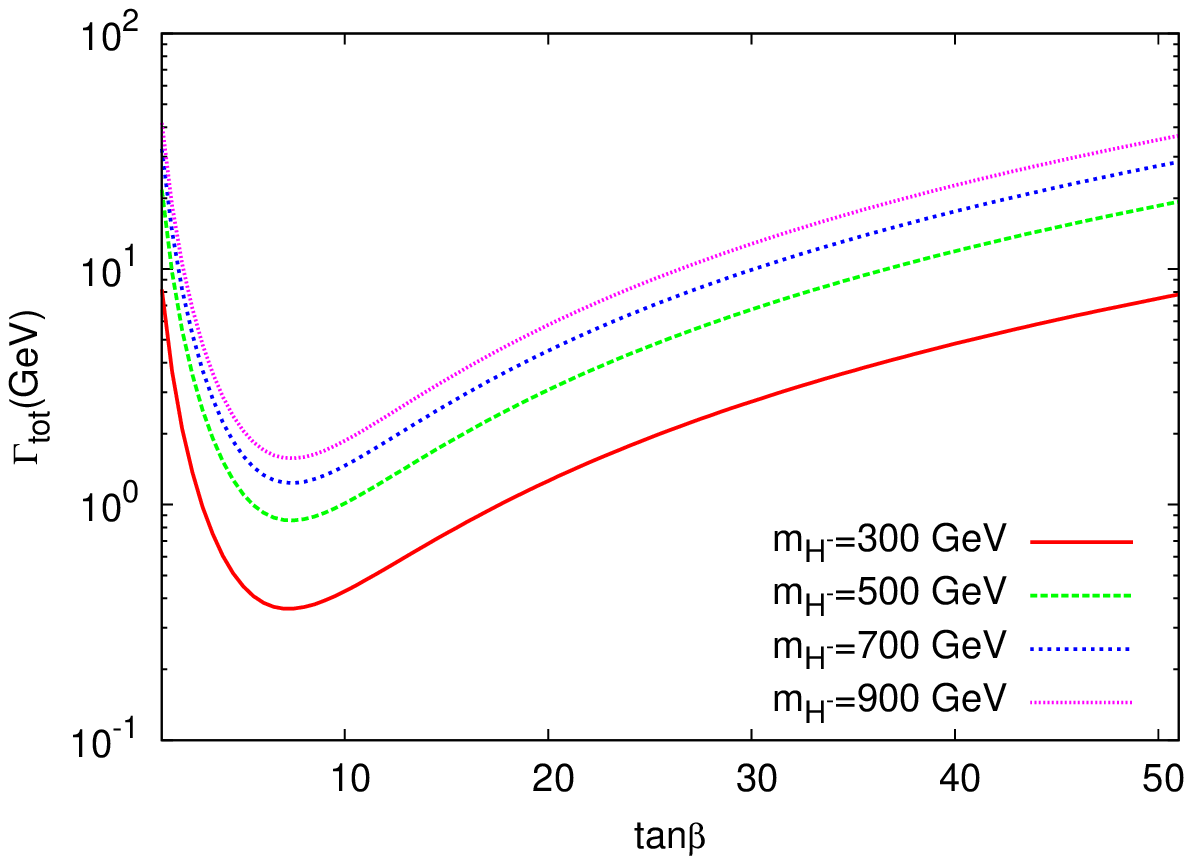}

\protect\caption{Decay width of $H^{\pm}$ boson depending on $\tan\beta$ for different
mass values of charged Higgs boson for parameter set PII of model
THDM-II. \label{fig:fig4}}
\end{figure}

The decay widths of the charged Higgs boson for the models THDM-I
and THDM-II are presented in Fig. \ref{fig:fig1}-\ref{fig:fig4}.
It is shown that the decay width for model THDM-I is larger than that
for THDM-II in the considered parameter space. The decay width increases
with the charged Higgs boson mass for parameter PI of THDM-I, while
it slightly changes depending on the mass $m_{H^{-}}>300$ GeV for
parameter PII of THDM-I, and it is almost constant for large $\tan\beta$.
In the model THDM-II, the decay width shows a minimum around $\tan\beta\approx7$.
As an example, using the parametrization PI (PII) of THDM-II, the
mass $m_{H^{-}}=300$ GeV and $\tan\beta=7$, the decay width is obtained
$\Gamma=5.0\times10^{0}$ ($3.5\times10^{-1}$) GeV as in Fig. \ref{fig:fig3}
(\ref{fig:fig4}), respectively.

\begin{figure}
\includegraphics[scale=0.8]{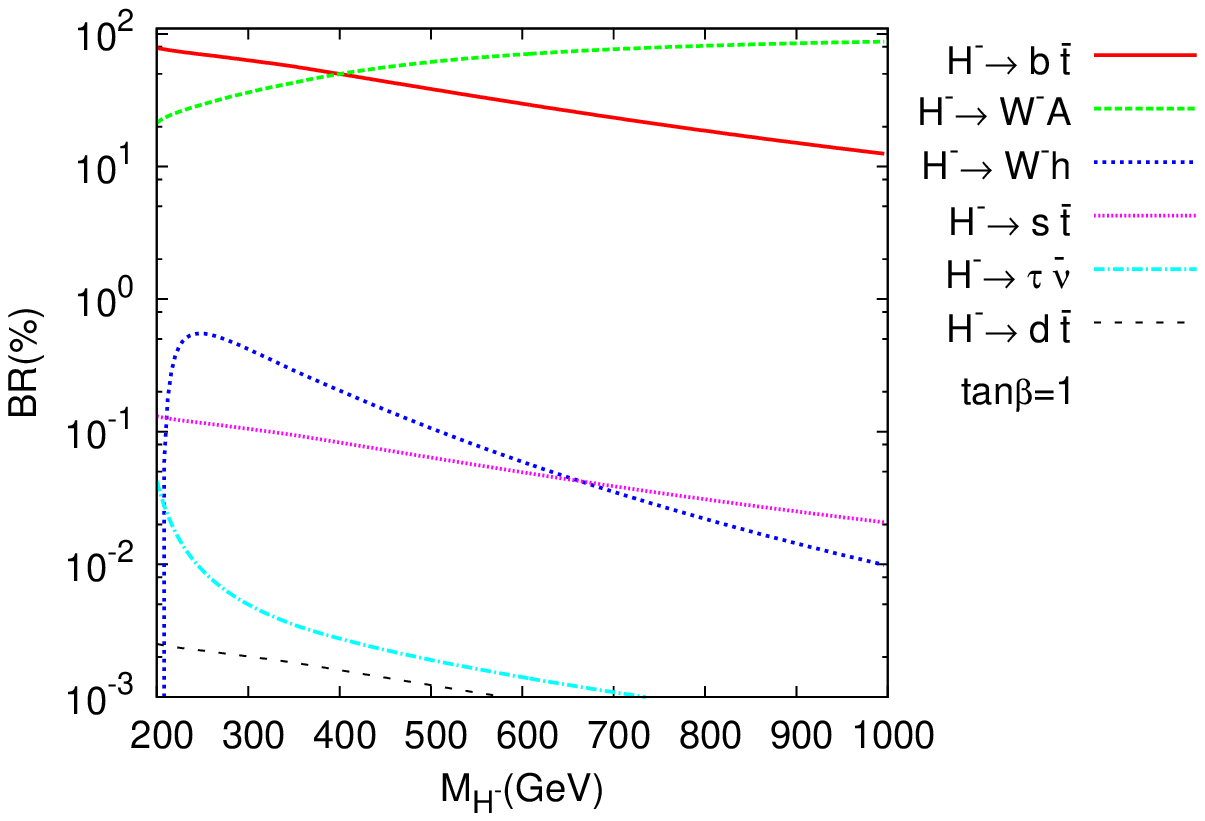}

\protect\caption{Branching ratios to different decay modes of charged Higgs boson depending
on $\tan\beta$ and parameter set PI of model THDM-I. \label{fig:fig5}}
\end{figure}

\begin{figure}
\includegraphics[scale=0.8]{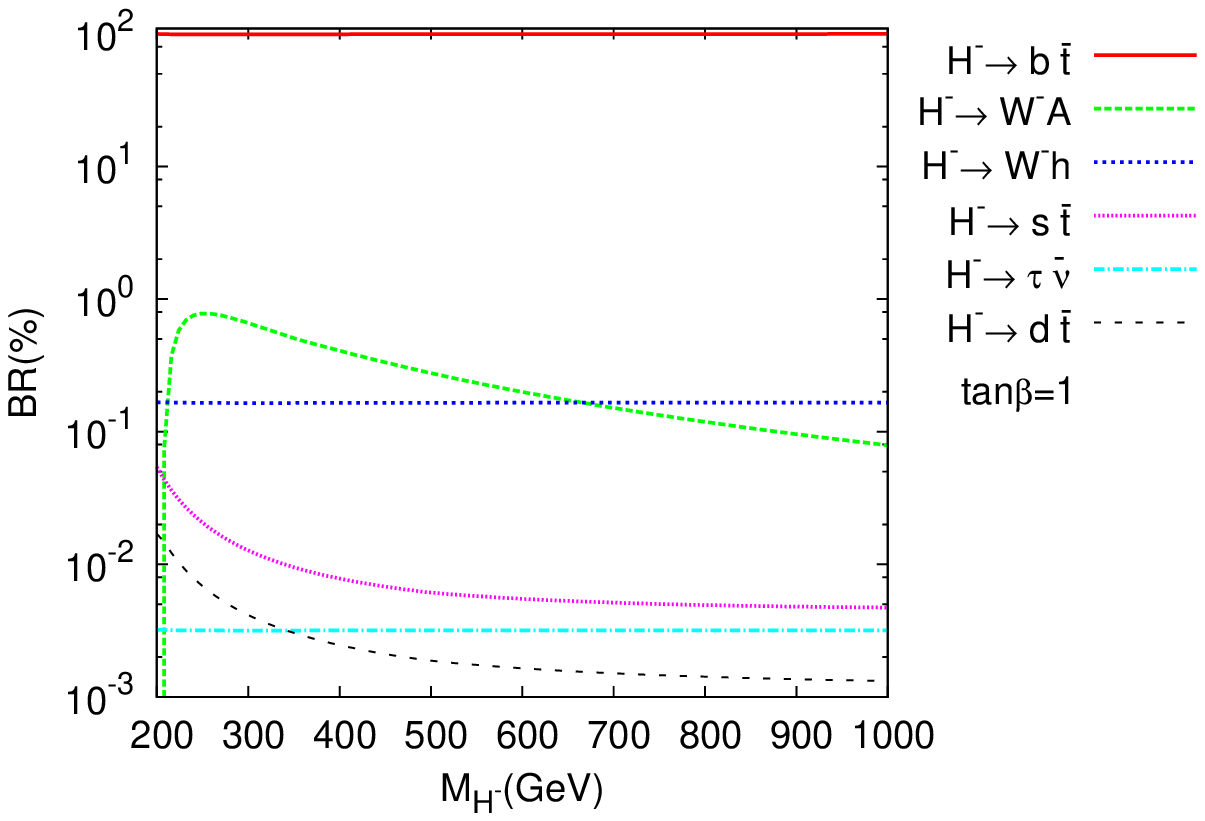}

\protect\caption{Branching ratios to different decay modes of charged Higgs boson depending
on $\tan\beta$ and parameter set PII of model THDM-I. \label{fig:fig6}}
\end{figure}

\begin{figure}
\includegraphics[scale=0.8]{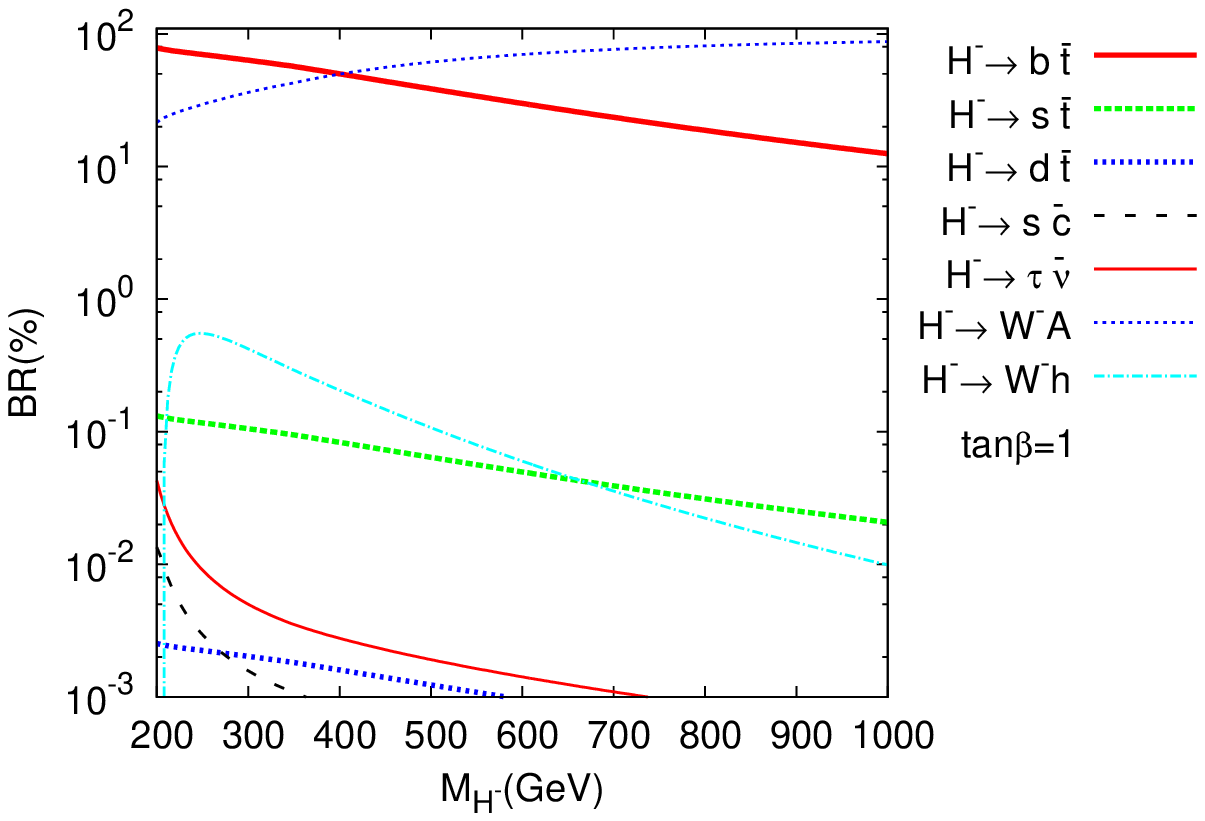}

\protect\caption{Branching ratios to different decay modes of charged Higgs boson depending
on its mass for $\tan\beta=1$ and parameter set PI of model THDM-II.
\label{fig:fig7}}
\end{figure}

\begin{figure}
\includegraphics[scale=0.8]{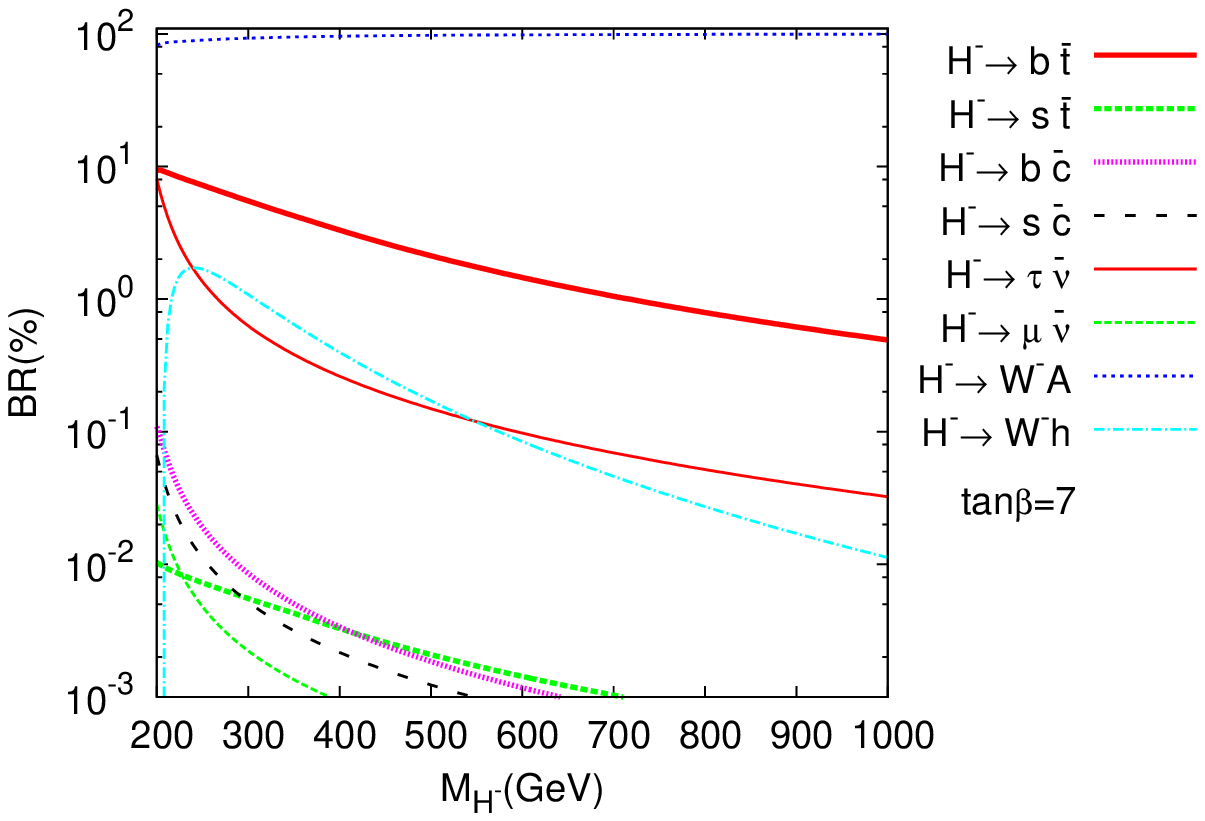}

\protect\caption{Branching ratios to different decay modes of charged Higgs boson depending
on its mass for $\tan\beta=7$ and parameter set PI of model THDM-II.
\label{fig:fig8}}
\end{figure}

\begin{figure}
\includegraphics[scale=0.8]{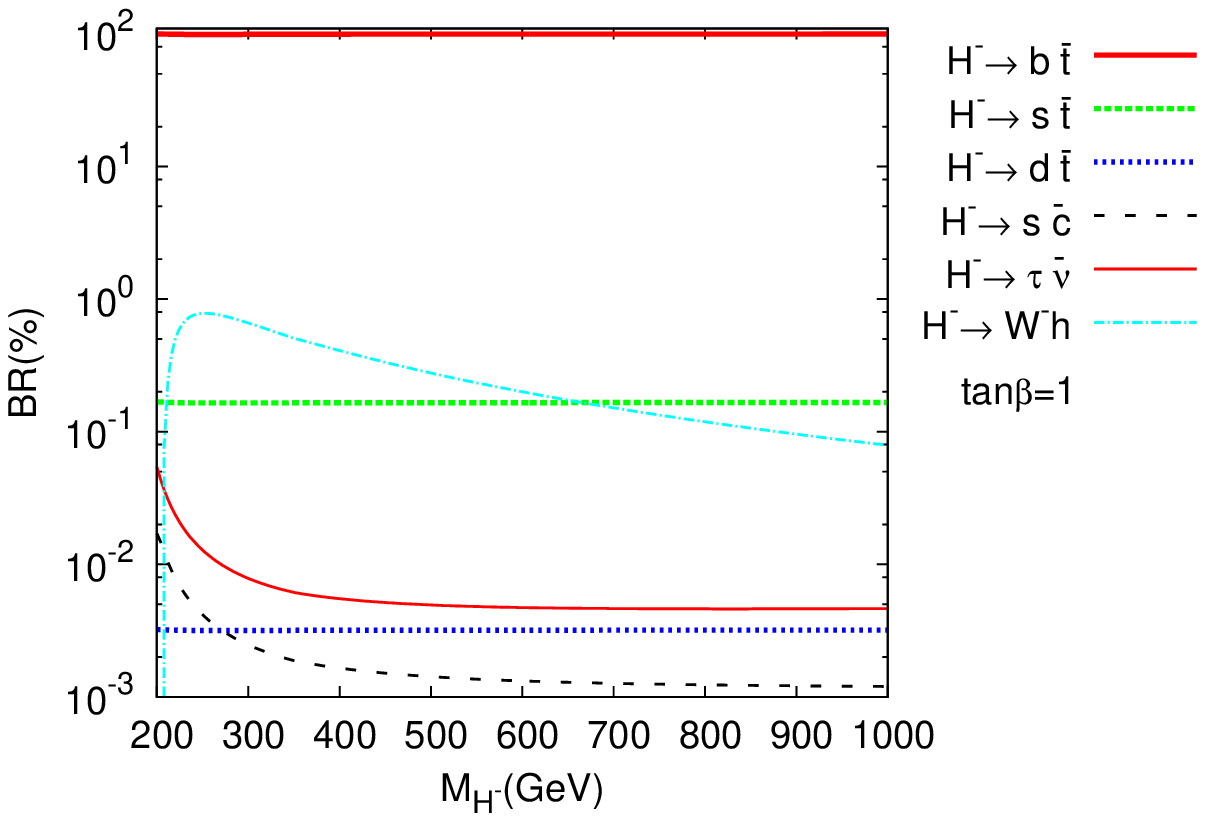}

\protect\caption{Branching ratios to different decay modes of charged Higgs boson depending
on its mass for $\tan\beta=1$ for parameter set PII and model THDM-II.
\label{fig:fig9}}
\end{figure}

\begin{figure}
\includegraphics[scale=0.8]{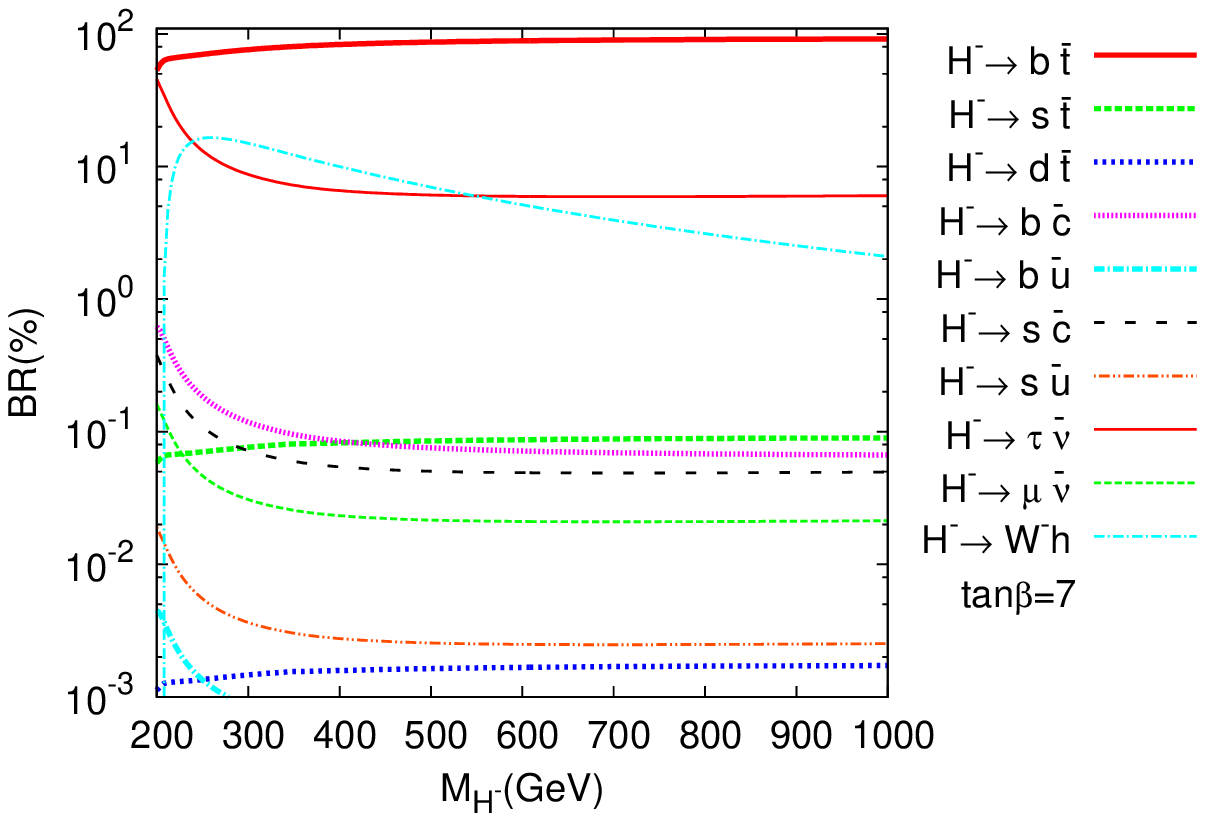}

\protect\caption{Branching ratios to different decay modes of charged Higgs boson depending
on its mass for $\tan\beta=7$ for parameter set PII and model THDM-II.
\label{fig:fig10}}
\end{figure}

For the parameter set PI and PII of THDM-I, the branching ratios of
charged Higgs boson into different decay channels are given in Fig.
\ref{fig:fig5}-\ref{fig:fig6}. For a mass value of $m_{A^{0}}>100$
GeV the branching into the channel $H^{-}\to\bar{t}b$ becomes dominant
as shown in Fig. \ref{fig:fig6}. The branching ratios are given in
Fig. \ref{fig:fig7}-\ref{fig:fig10} for the parameter set PI and
PII of THDM-II.

\section{Production cross section at FCC-hh collider}

\begin{figure}
\includegraphics[scale=0.8]{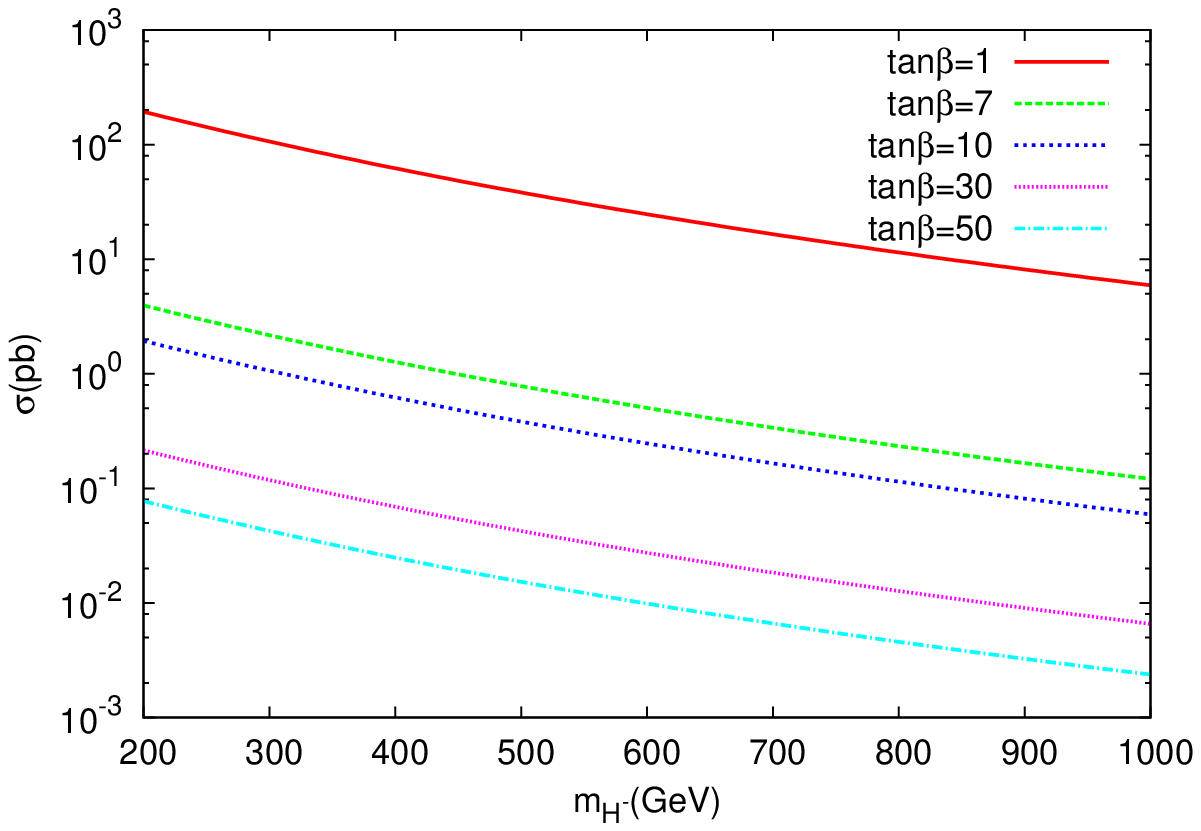}

\protect\caption{Cross section for charged Higgs boson production within the THDM-I
at FCC-hh collider. \label{fig:fig11}}
\end{figure}

\begin{figure}
\includegraphics[scale=0.8]{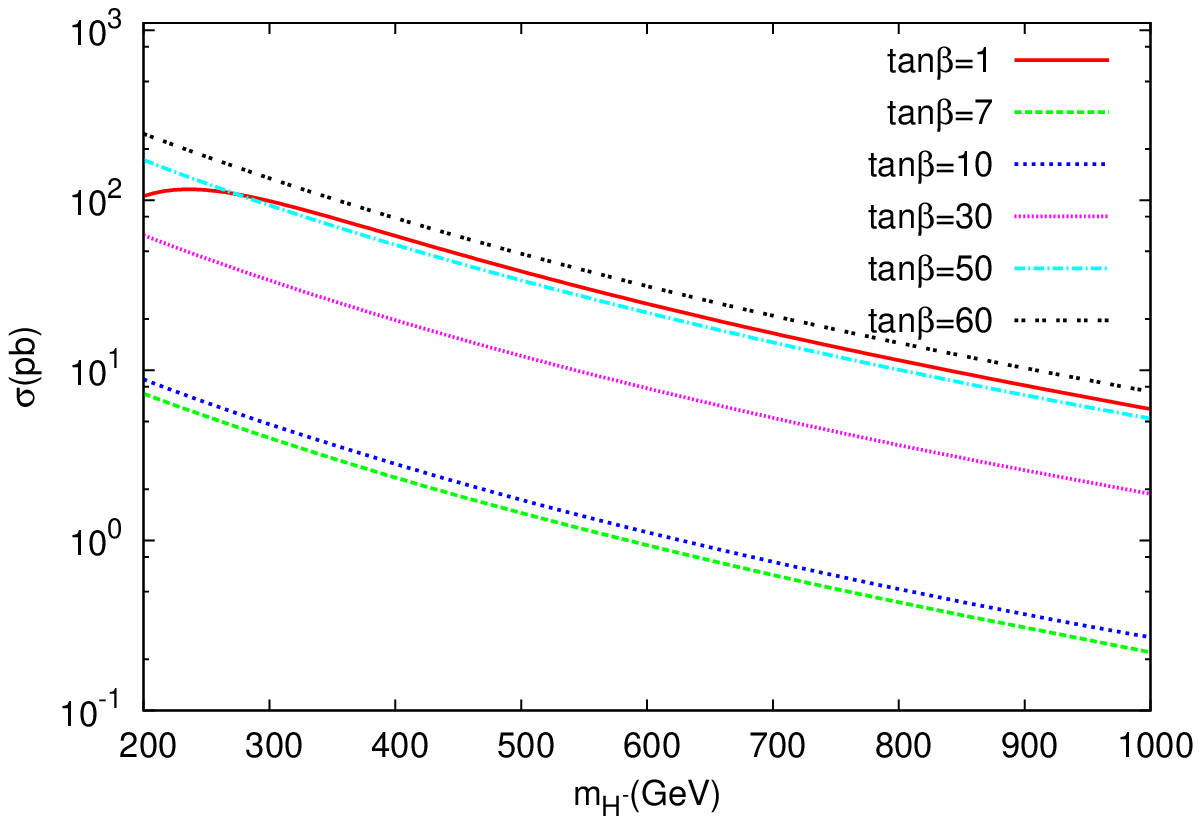}

\protect\caption{Cross section for charged Higgs boson production within the THDM-II
at FCC-hh collider. \label{fig:fig12}}
\end{figure}

The ongoing searches at the LHC rely on specific production and decay
mechanism that occupy only a part of the complete model parameter
space. The cross sections for the single production of charged Higgs
boson through the process $pp\to tH^{-}+X$ within the THDM-I and
THDM-II are presented in Fig. \ref{fig:fig11} and \ref{fig:fig12}.
The characteristics of the cross sections depending on the $\tan\beta$
can be seen from these figures as expected from the $H^{-}f_{i}\bar{f}_{j}$
interaction vertices. The cross sections have a minimum around $\tan\beta\approx7$
for THDM-II while it decreases with the increasing values of $\tan\beta$
for THDM-I. There exist large values of cross section $\sigma\simeq37.5$
pb for $\tan\beta=1$, however it is $0.8$ pb ($1.5$ pb) for $\tan\beta=7$
within the model THDM-I (THDM-II), respectively. For large values
of $\tan\beta$ the cross section decreases for THDM-I, while it is
in the same level (comparing the cross sections at $\tan\beta=1$
and $\tan\beta\simeq50$) for THDM-II.

\begin{figure}
\includegraphics[scale=0.8]{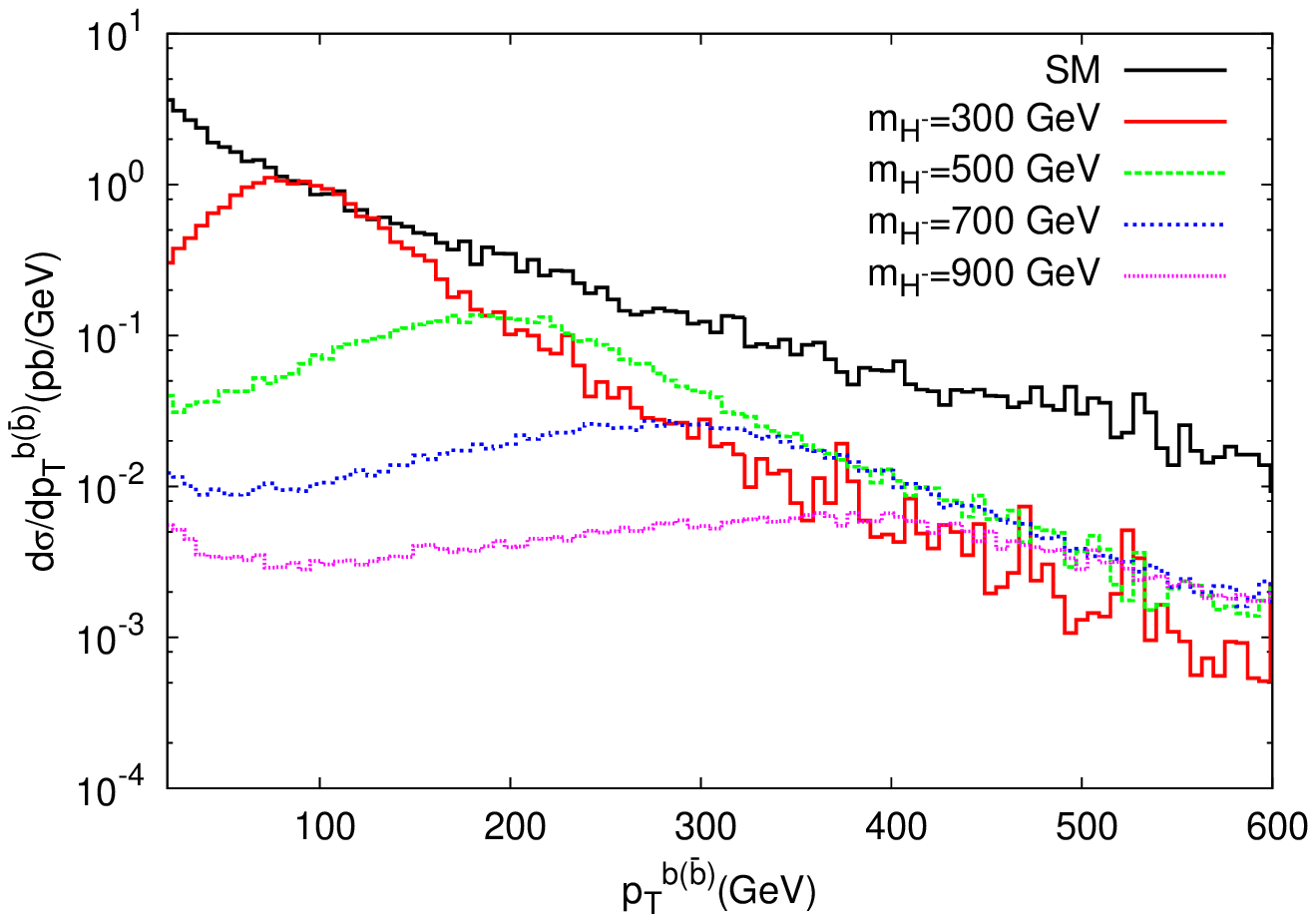}

\protect\caption{Transverse momentum distributions of $b(\bar{b})$ quarks for set
PI of THDM-I. \label{fig:fig13}}
\end{figure}

\begin{figure}
\includegraphics[scale=0.8]{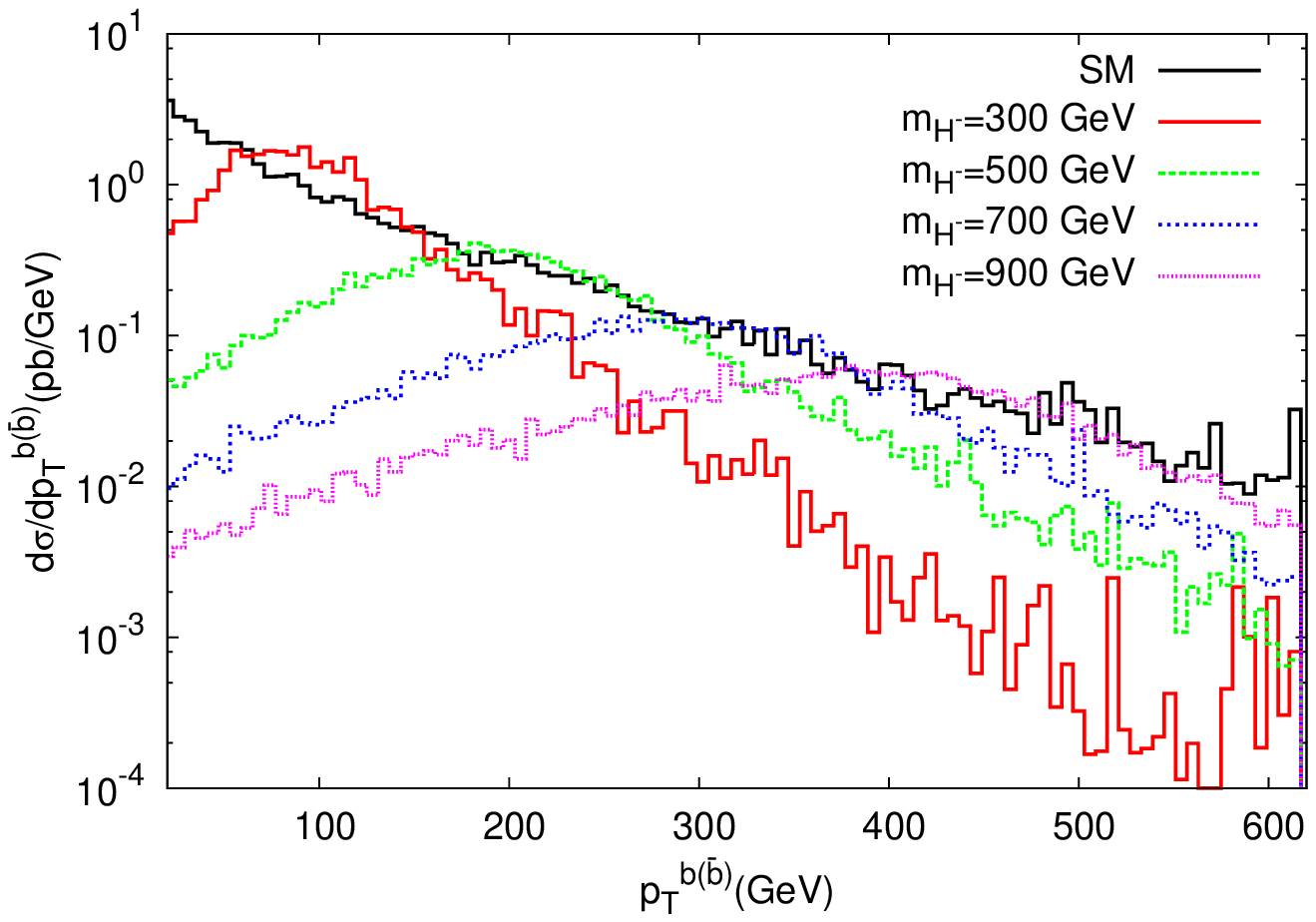}

\protect\caption{Transverse momentum distributions of $b(\bar{b})$ quarks for parameter
PII of THDM-I. \label{fig:fig14}}
\end{figure}

\begin{figure}
\includegraphics[scale=0.8]{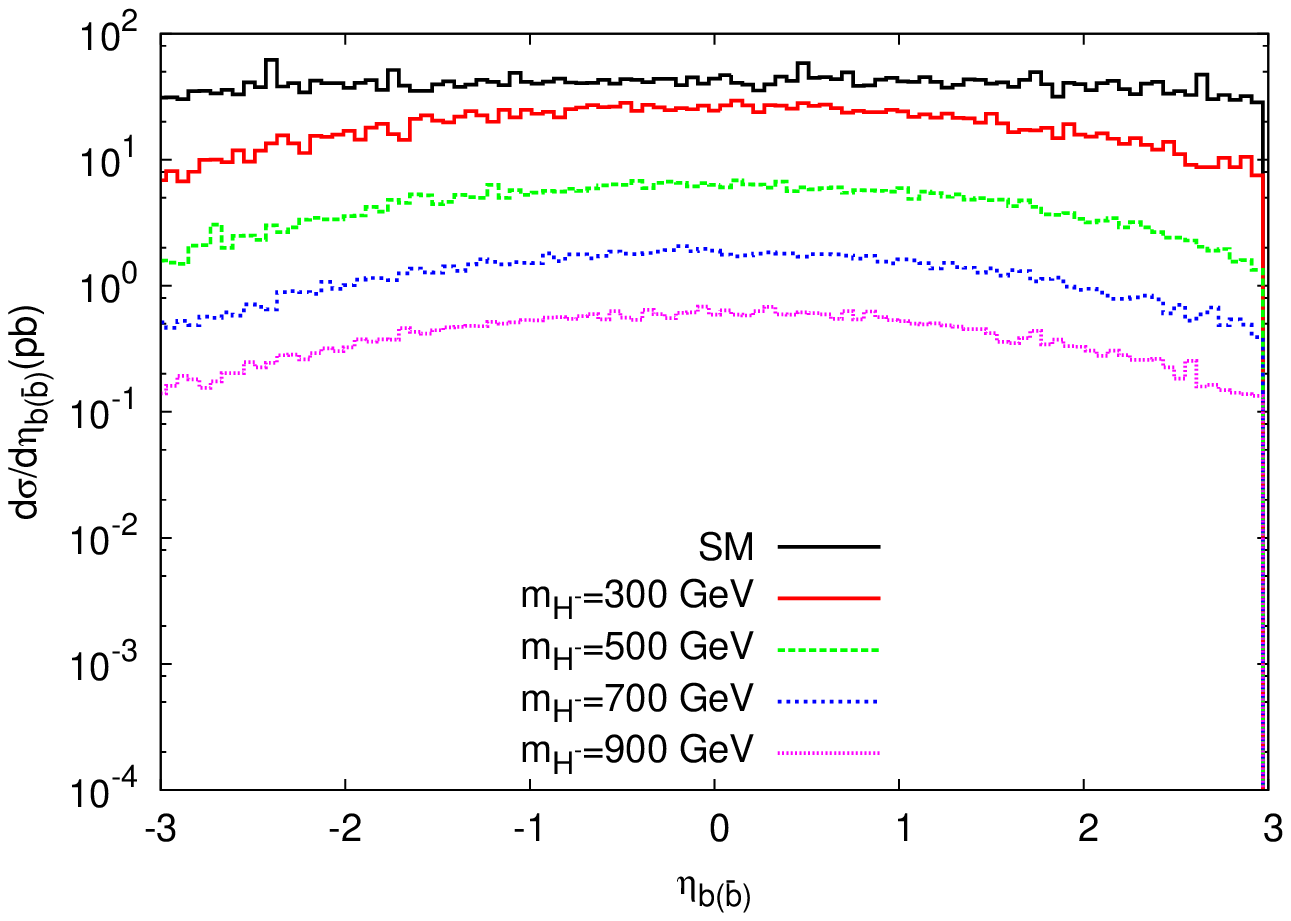}

\protect\caption{Pseudorapidity distributions of $b(\bar{b})$ quarks for set PI of
THDM-I. \label{fig:fig15}}
\end{figure}

\begin{figure}
\includegraphics[scale=0.8]{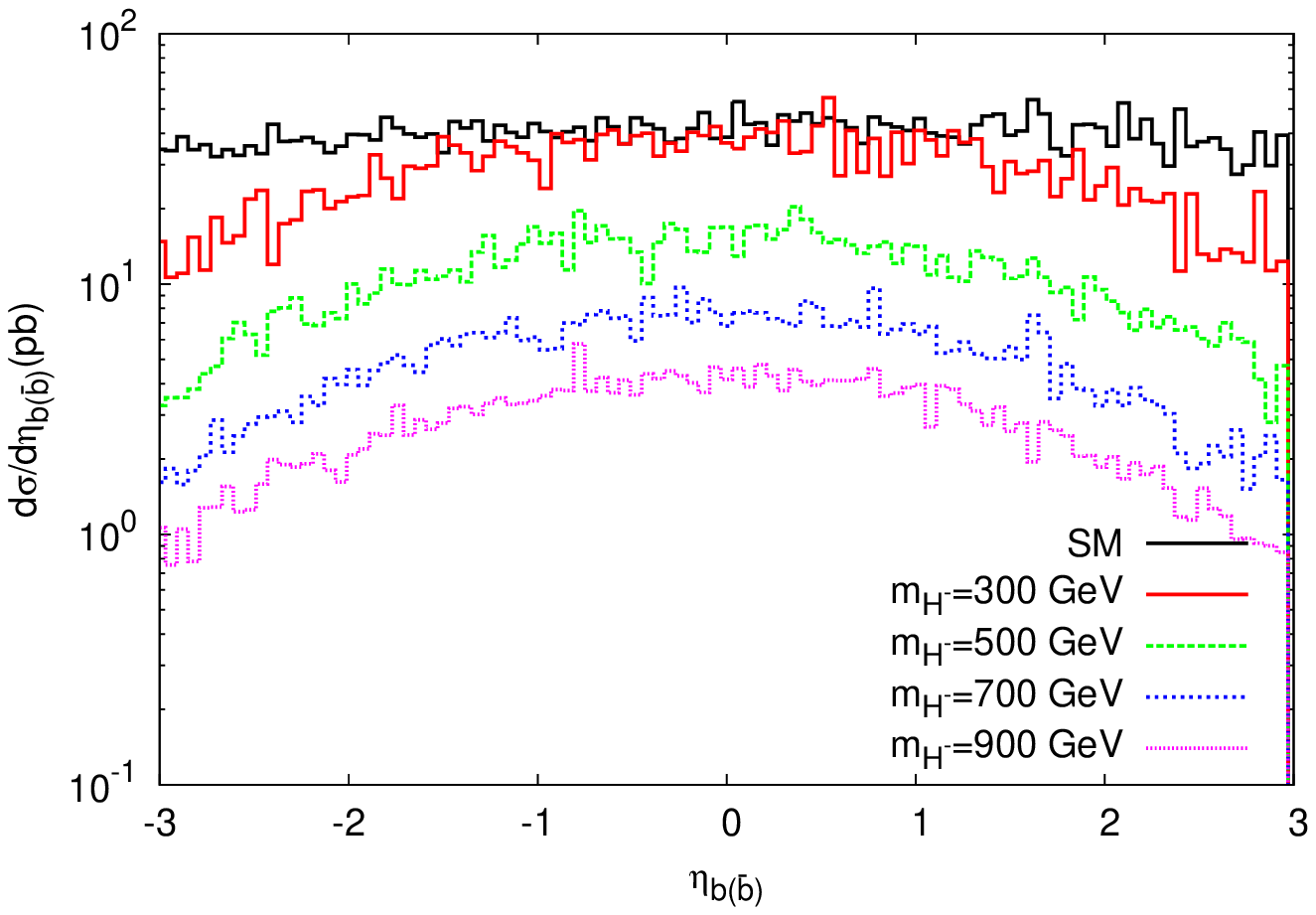}

\protect\caption{Pseudorapidity distributions of $b(\bar{b})$ quarks for parameter
PII of THDM-I. \label{fig:fig16}}
\end{figure}

\begin{figure}
\includegraphics[scale=0.8]{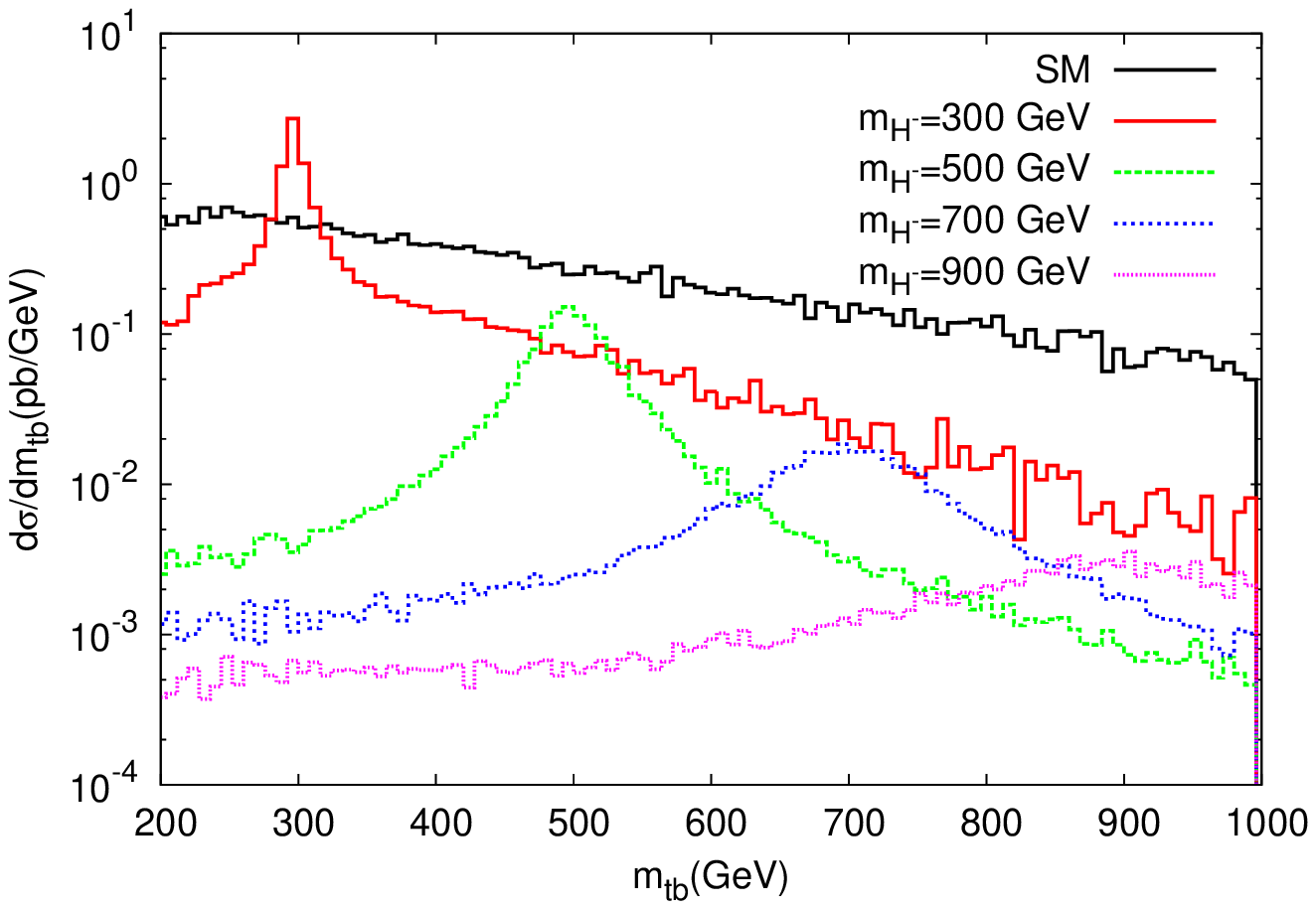}

\protect\caption{Invariant mass distributions of $b(\bar{b})$ quarks for set PI of
THDM-I. \label{fig:fig17}}
\end{figure}

\begin{figure}
\includegraphics[scale=0.8]{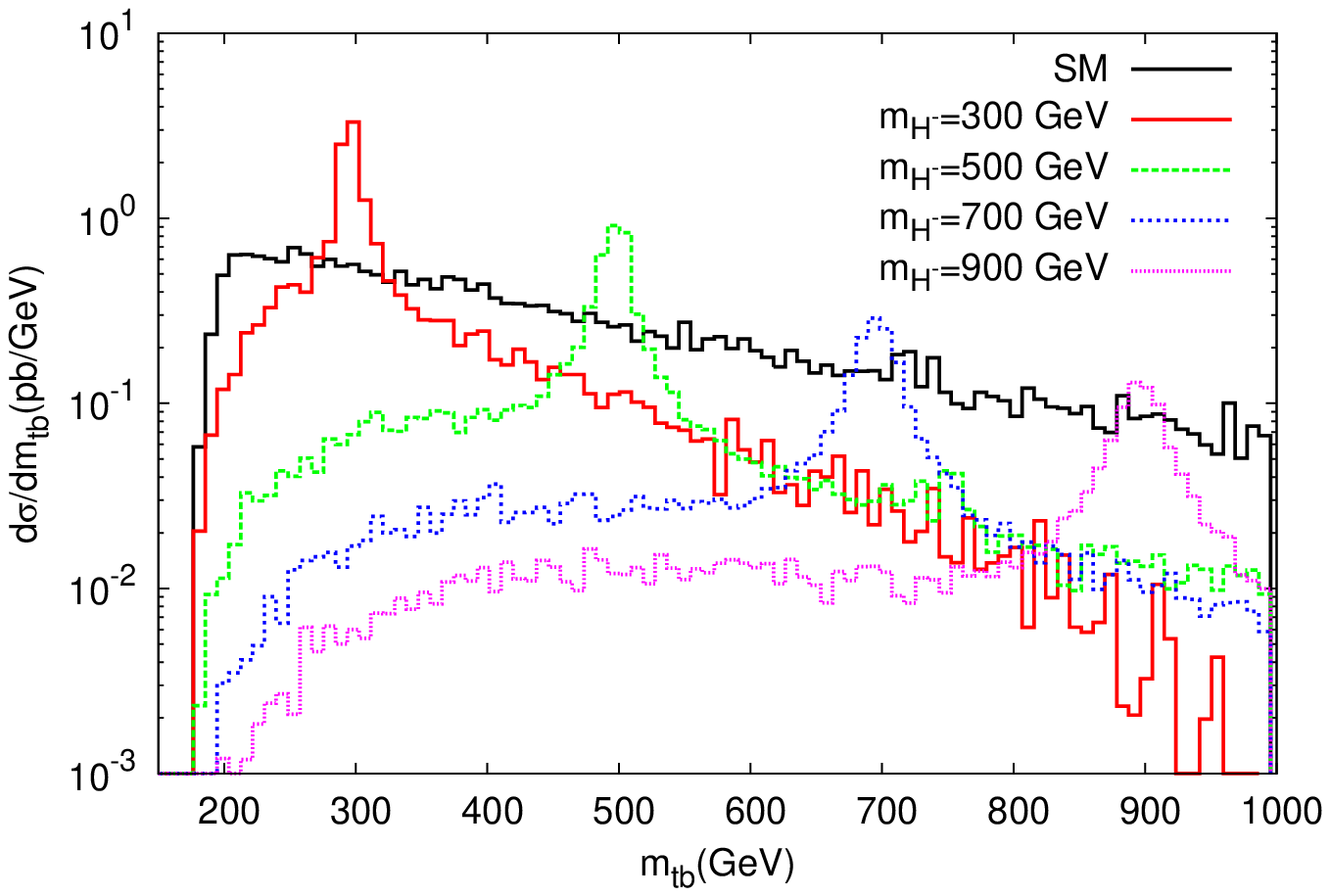}

\protect\caption{Invariant mass distributions of $b(\bar{b})$ quarks for parameter
PII of THDM-I. \label{fig:fig18}}
\end{figure}

In order to examine the kinematical distributions of associated $b(\bar{b})$
quark, the transverse momentum distributions of $b(\bar{b})$ quarks
for the process $pp\to t\bar{t}b(\bar{b})+X$ are presented in Fig.
\ref{fig:fig13} and \ref{fig:fig14} for the parameter set PI and
PII of THDM-I, respectively. Fig. \ref{fig:fig15} and \ref{fig:fig16}
show the pseudorapidity distributions of $b(\bar{b})$ quarks for
parameter set PI and PII of THDM-I, respectively. The invariant mass
distributions $m_{tb}$ ($t$ and $b(\bar{b})$ quark in the final
state) are presented in Fig. \ref{fig:fig17} and \ref{fig:fig18}
for parameter PI and PII of THDM-I, respectively.

\begin{figure}
\includegraphics[scale=0.8]{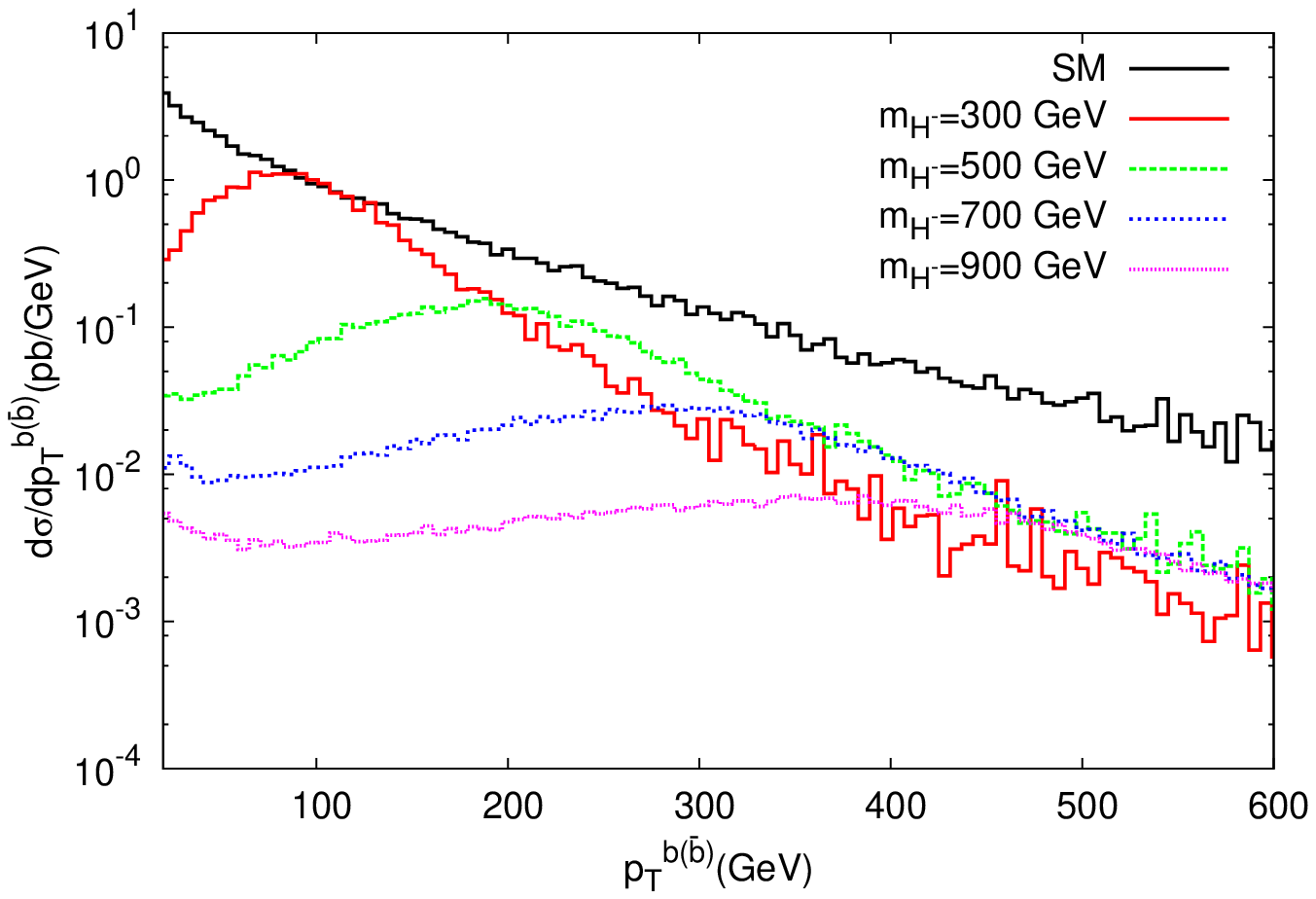}

\protect\caption{Transverse momentum distributions of $b(\bar{b})$ quarks for set
PI of THDM-II. \label{fig:fig19}}
\end{figure}

\begin{figure}
\includegraphics[scale=0.8]{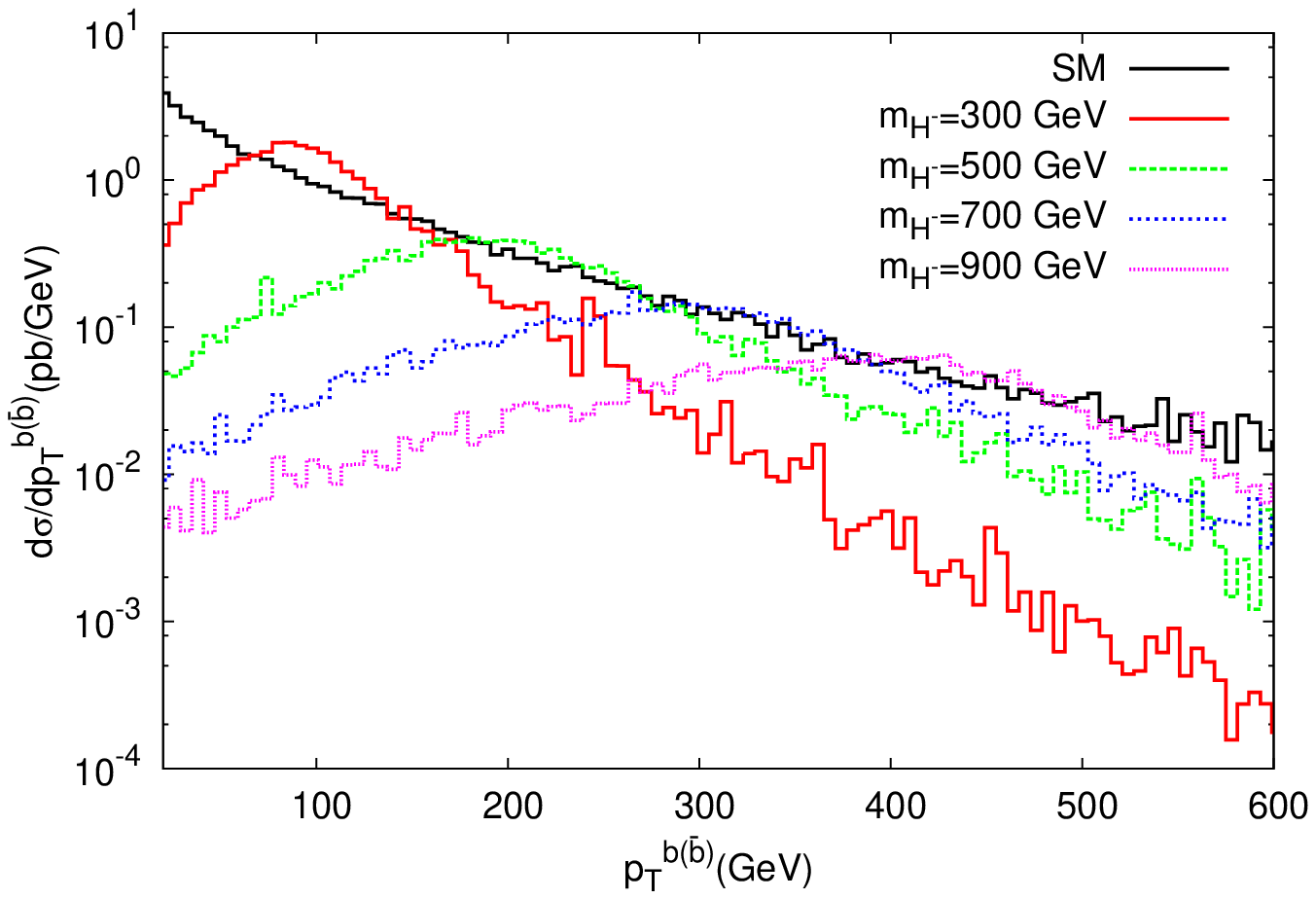}

\protect\caption{Transverse momentum distributions of $b(\bar{b})$ quarks for parameter
PII of THDM-II. \label{fig:fig20}}
\end{figure}

\begin{figure}
\includegraphics[scale=0.8]{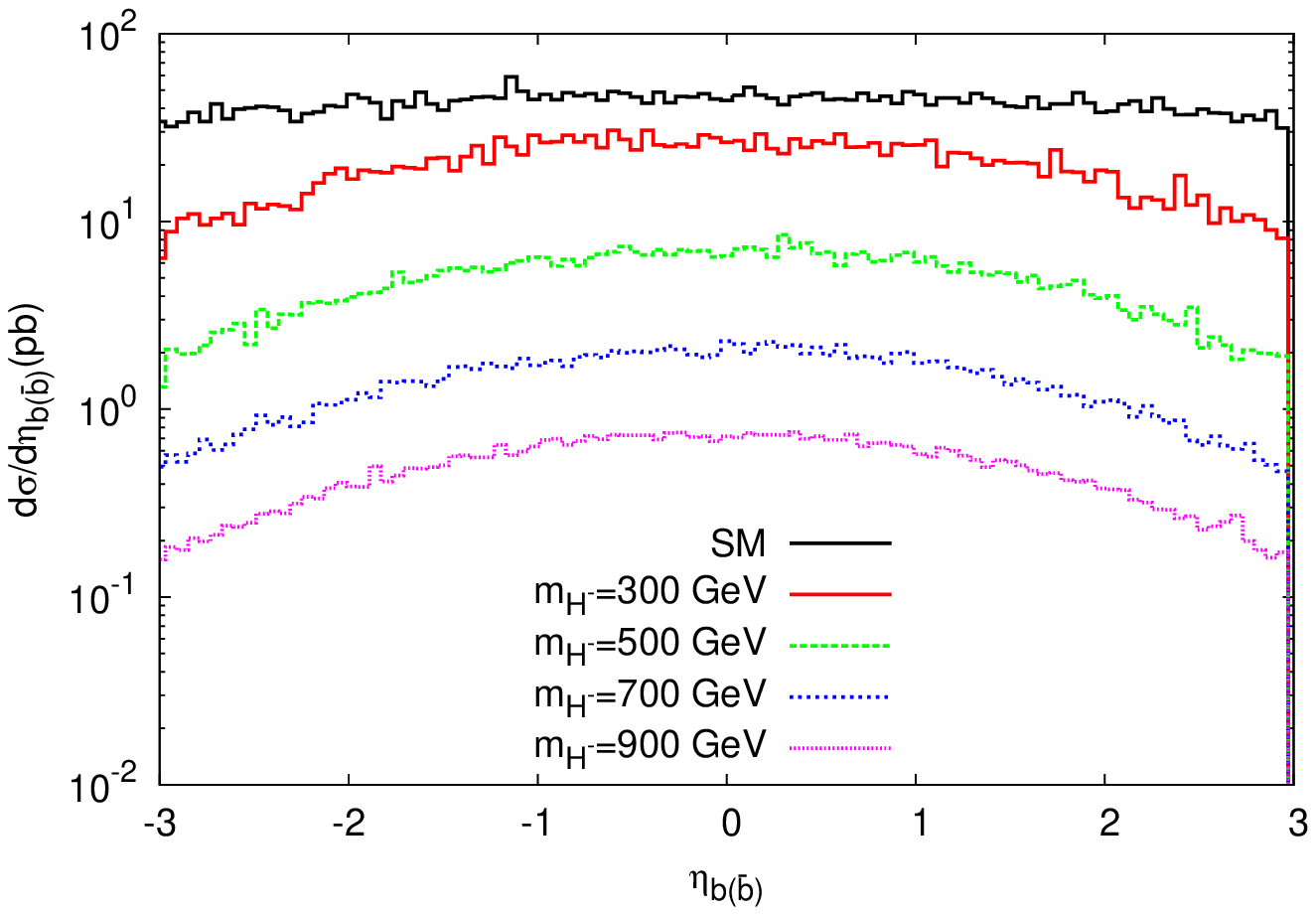}

\protect\caption{Pseudorapidity distributions of $b(\bar{b})$ quarks for set PI of
THDM-II. \label{fig:fig21}}
\end{figure}

\begin{figure}
\includegraphics[scale=0.8]{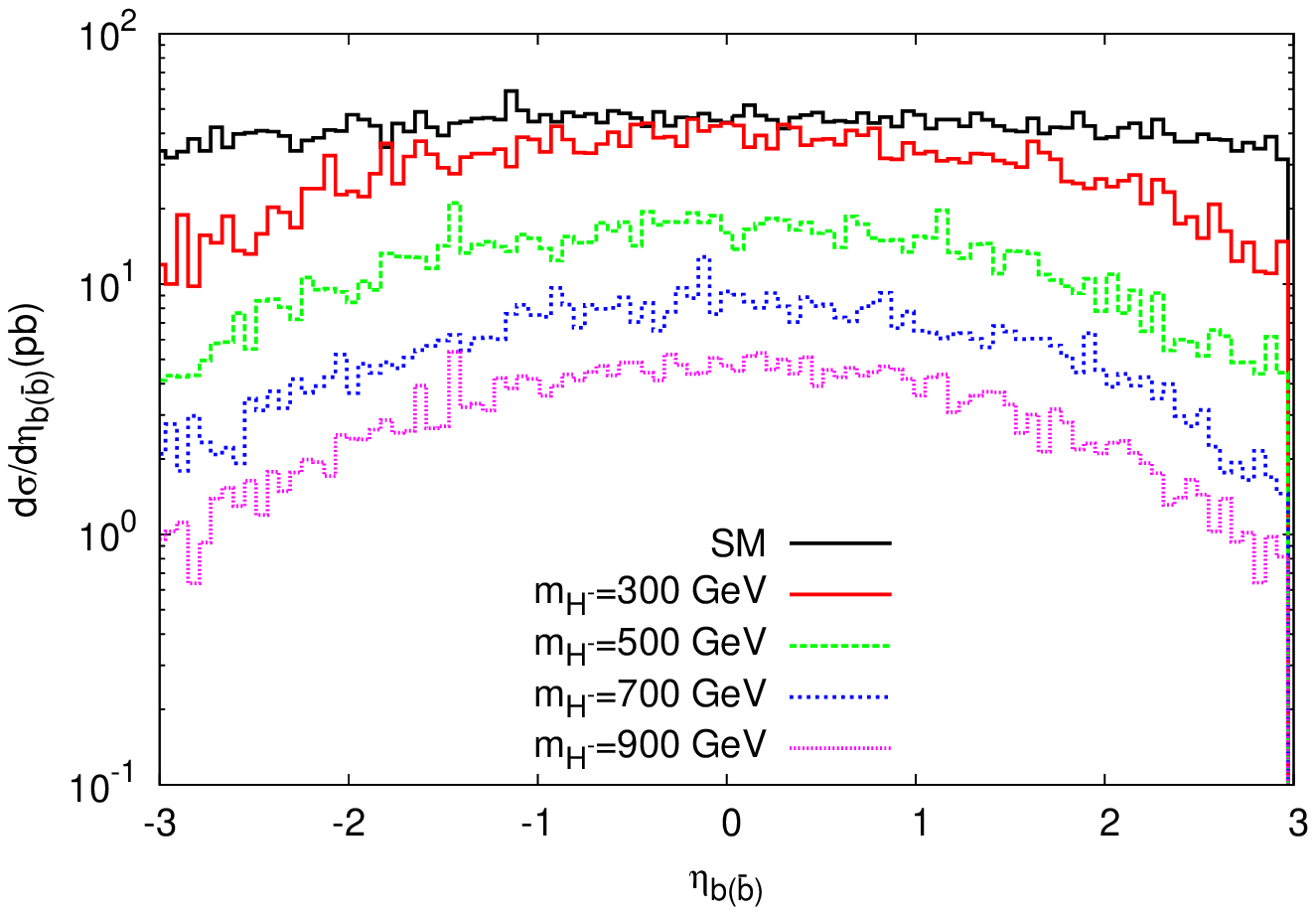}

\protect\caption{Pseudorapidity distributions of $b(\bar{b})$ quarks for parameter
PII of THDM-II. \label{fig:fig22}}
\end{figure}

\begin{figure}
\includegraphics[scale=0.8]{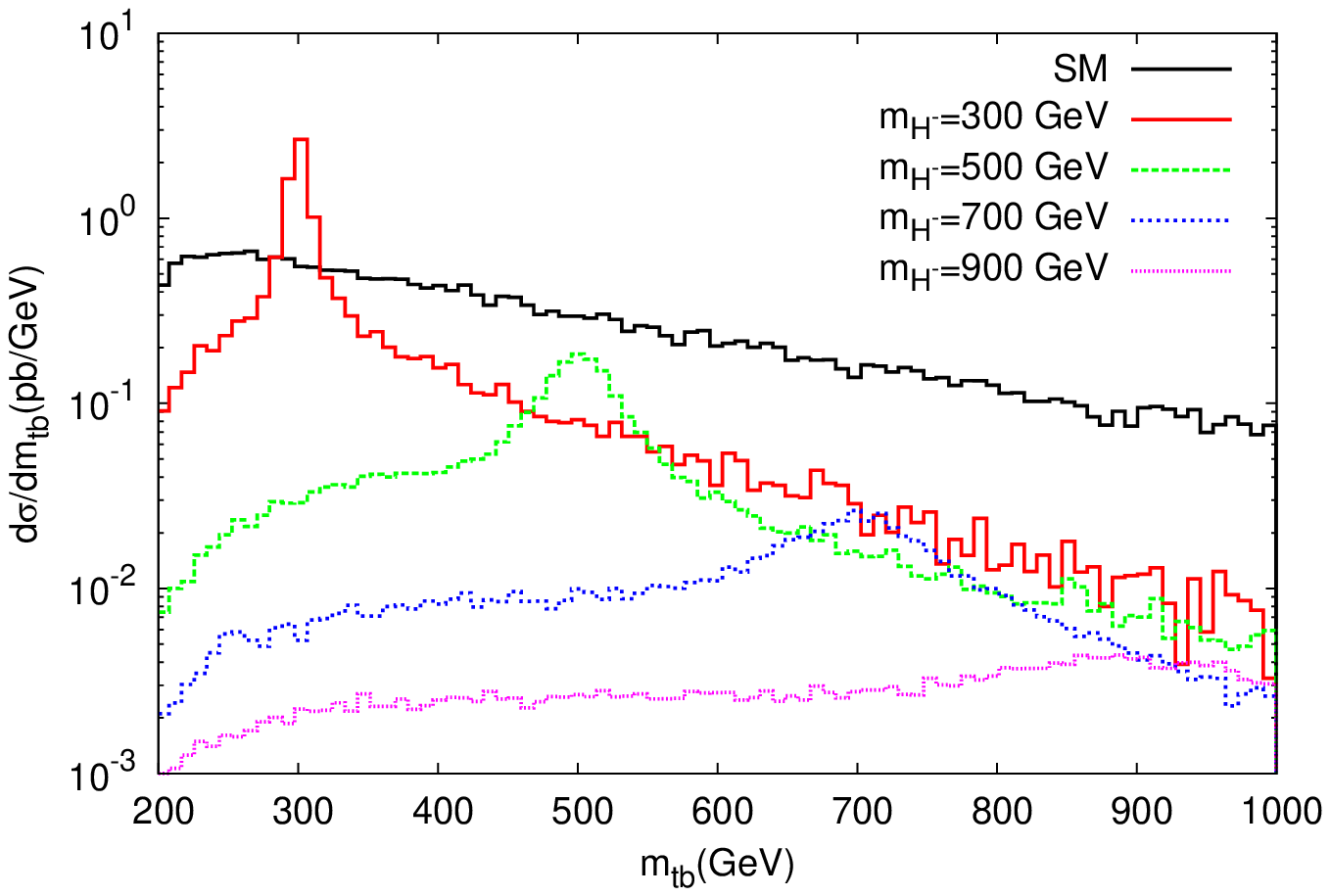}

\protect\caption{Invariant mass distributions of $b(\bar{b})$ quarks for set PI of
THDM-II. \label{fig:fig23}}
\end{figure}

\begin{figure}
\includegraphics[scale=0.8]{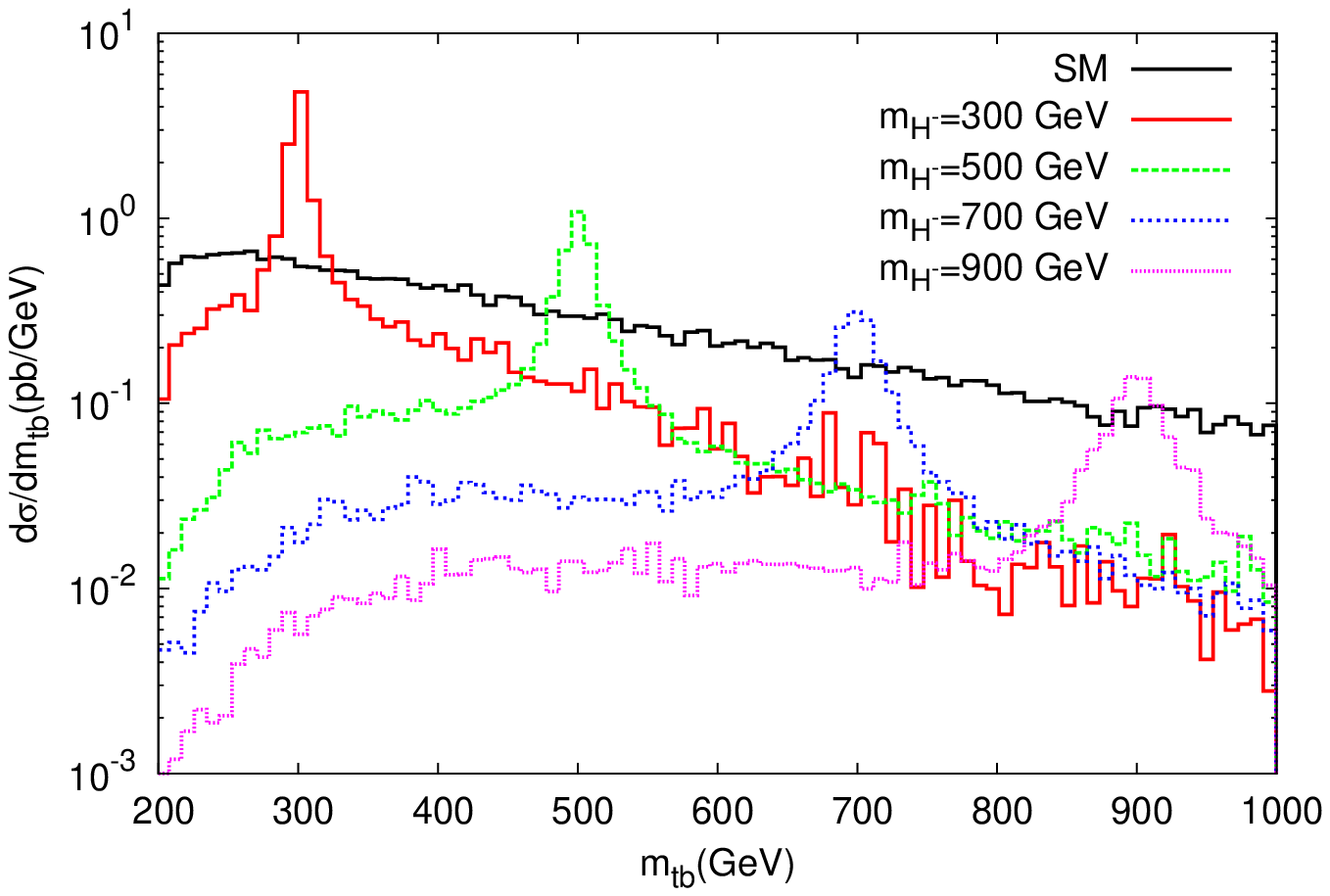}

\protect\caption{Invariant mass distributions of $b(\bar{b})$ quarks for parameter
PII of THDM-II. \label{fig:fig24}}
\end{figure}

Transverse momentum distributions of $b(\bar{b})$ quarks are presented
in Fig. \ref{fig:fig19} and \ref{fig:fig20} for parameter set PI
and PII of THDM-II, respectively. Fig. \ref{fig:fig21} and \ref{fig:fig22}
show the pseudorapidity distributions of $b(\bar{b})$ quarks for
parameter set PI and PII of THDM-II, respectively. The invariant mass
distributions $m_{tb}$ (top and bottom quark in the final state)
are presented in Fig. \ref{fig:fig23} and \ref{fig:fig24} for parameter
PI and PII of THDM-II, respectively.

The cross sections for the background processes $pp\to t\bar{t}b(\bar{b})+X$,
$pp\to t\bar{t}c(\bar{c})+X$ and $pp\to t\bar{t}j+X$ are presented
in Table \ref{tab:table1} for the center of mass energy $\sqrt{s}=100$
TeV of FCC-hh collider. In the study we consider the background due
to one leading $b$-jet. Other $b$-jets originating from top quark
decays will have different kinematical distributions from the top-antitop
associated leading one. In addition to the basic kinematical cuts
$p_{T}>20$ GeV and $|\eta|<2.5$ an invariant mass cut $|m_{tb}-m_{H^{\pm}}|\leq0.1m_{H^{\pm}}$
is applied for the analysis. For the calculation of background cross
section $\Delta\sigma_{B}$ within the invariant mass interval $|m_{tb}-m_{H^{\pm}}|\leq0.1m_{H^{\pm}}$
, we assume the efficiency of $b$-tagging to be $\epsilon_{b}=50\%$
and the rejection ratios to be $10\%$ for $c(\bar{c})$ quark jets
and $1\%$ for light quark jets since they are assumed to be mistagged
as $b$-jets.

\begin{table}
\protect\caption{The cross sections (in pb) for the background processes $pp\to t\bar{t}b(\bar{b})+X$,
$pp\to t\bar{t}c(\bar{c})+X$ and $pp\to t\bar{t}j+X$ calculated
in the invariant mass range $|m_{tb}-m_{H^{\pm}}|\leq0.1m_{H^{\pm}}$
at the center of mass energy $\sqrt{s}=100$ TeV. \label{tab:table1}}

\begin{tabular}{|c|c|c|c|c|}
\hline 
$m_{tb}$ (GeV) & $pp\to t\bar{t}b(\bar{b})+X$ & $pp\to t\bar{t}c(\bar{c})+X$ & $pp\to t\bar{t}j+X$ & $\Delta\sigma_{B}(pb)$\tabularnewline
\hline 
\hline 
$300\pm30$ & $2.99\times10^{0}$ & $3.65\times10^{0}$ & $2.51\times10^{2}$ & $4.37\times10^{0}$\tabularnewline
\hline 
$500\pm50$ & $2.88\times10^{0}$ & $3.62\times10^{0}$ & $1.23\times10^{2}$ & $3.03\times10^{0}$\tabularnewline
\hline 
$700\pm70$ & $1.89\times10^{0}$ & $2.40\times10^{0}$ & $8.04\times10^{1}$ & $1.99\times10^{0}$\tabularnewline
\hline 
$900\pm90$ & $1.34\times10^{0}$ & $1.86\times10^{0}$ & $5.47\times10^{1}$ & $1.40\times10^{0}$\tabularnewline
\hline 
\end{tabular}
\end{table}

In Tables \ref{tab:table2}-\ref{tab:table5} we present the signal
cross sections $\Delta\sigma_{S}$ within the invariant mass interval
$|m_{tb}-m_{H^{\pm}}|\leq0.1m_{H^{\pm}}$ , for single production
of charged Higgs boson in the model framework of THDM-I and THDM-II
(for the parametrizations PI and PII) at FCC-hh collider with $\sqrt{s}=100$
TeV.

\begin{table}
\protect\caption{The cross sections for the signal process $pp\to t\bar{t}b(\bar{b})+X$
within the THDM-I and parametrization PI calculated in the invariant
mass range $|m_{tb}-m_{H^{\pm}}|\leq0.1m_{H^{\pm}}$ at the center
of mass energy $\sqrt{s}=100$ TeV. \label{tab:table2}}

\begin{tabular}{|c|c|c|c|c|}
\hline 
THDM-I and PI & \multicolumn{4}{c|}{$\Delta\sigma_{S}(pb)$}\tabularnewline
\hline 
$m_{tb}$ (GeV) & $\tan\beta=1$ & $\tan\beta=7$ & $\tan\beta=30$ & $\tan\beta=50$\tabularnewline
\hline 
\hline 
$300\pm30$ & $1.24\times10^{1}$ & $1.30\times10^{-2}$ & $4.09\times10^{-5}$ & $5.02\times10^{-6}$\tabularnewline
\hline 
$500\pm50$ & $2.29\times10^{0}$ & $1.97\times10^{-3}$ & $5.02\times10^{-6}$ & $5.96\times10^{-7}$\tabularnewline
\hline 
$700\pm70$ & $5.52\times10^{-1}$ & $4.60\times10^{-4}$ & $8.99\times10^{-7}$ & $1.19\times10^{-7}$\tabularnewline
\hline 
$900\pm90$ & $1.13\times10^{-1}$ & $6.59\times10^{-5}$ & $2.43\times10^{-7}$ & $2.54\times10^{-8}$\tabularnewline
\hline 
\end{tabular}
\end{table}

\begin{table}
\protect\caption{The same as \ref{tab:table2}, but for the THDM-I and PII. \label{tab:table3}}

\begin{tabular}{|c|c|c|c|c|}
\hline 
THDM-I and PII & \multicolumn{4}{c|}{$\Delta\sigma_{S}(pb)$}\tabularnewline
\hline 
$m_{tb}$ (GeV) & $\tan\beta=1$ & $\tan\beta=7$ & $\tan\beta=30$ & $\tan\beta=50$\tabularnewline
\hline 
\hline 
$300\pm30$ & $1.88\times10^{1}$ & $7.16\times10^{-2}$ & $4.21\times10^{-4}$ & $7.57\times10^{-5}$\tabularnewline
\hline 
$500\pm50$ & $8.33\times10^{0}$ & $5.56\times10^{-2}$ & $1.94\times10^{-4}$ & $5.03\times10^{-5}$\tabularnewline
\hline 
$700\pm70$ & $3.41\times10^{0}$ & $1.84\times10^{-2}$ & $1.64\times10^{-4}$ & $3.02\times10^{-5}$\tabularnewline
\hline 
$900\pm90$ & $1.59\times10^{0}$ & $1.72\times10^{-2}$ & $9.28\times10^{-5}$ & $6.32\times10^{-6}$\tabularnewline
\hline 
\end{tabular}
\end{table}

\begin{table}
\protect\caption{The same as \ref{tab:table2}, but for the THDM-II and PI. \label{tab:table4}}

\begin{tabular}{|c|c|c|c|c|}
\hline 
THDM-II and PI & \multicolumn{4}{c|}{$\Delta\sigma_{S}(pb)$}\tabularnewline
\hline 
$m_{tb}$ (GeV) & $\tan\beta=1$ & $\tan\beta=7$ & $\tan\beta=30$ & $\tan\beta=50$\tabularnewline
\hline 
\hline 
$300\pm30$ & $1.22\times10^{1}$ & $3.57\times10^{-2}$ & $5.22\times10^{-1}$ & $7.24\times10^{0}$\tabularnewline
\hline 
$500\pm50$ & $2.45\times10^{0}$ & $5.26\times10^{-3}$ & $3.48\times10^{-1}$ & $2.02\times10^{0}$\tabularnewline
\hline 
$700\pm70$ & $3.90\times10^{-1}$ & $9.37\times10^{-4}$ & $6.08\times10^{-2}$ & $2.97\times10^{-1}$\tabularnewline
\hline 
$900\pm90$ & $1.00\times10^{-1}$ & $1.86\times10^{-4}$ & $1.24\times10^{-2}$ & $8.59\times10^{-2}$\tabularnewline
\hline 
\end{tabular}
\end{table}

\begin{table}
\protect\caption{The same as \ref{tab:table2}, but for the THDM-II and PII. \label{tab:table5}}

\begin{tabular}{|c|c|c|c|c|}
\hline 
THDM-II and PII & \multicolumn{4}{c|}{$\Delta\sigma_{S}(pb)$}\tabularnewline
\hline 
$m_{tb}$ (GeV) & $\tan\beta=1$ & $\tan\beta=7$ & $\tan\beta=30$ & $\tan\beta=50$\tabularnewline
\hline 
\hline 
$300\pm30$ & $1.70\times10^{1}$ & $2.12\times10^{-1}$ & $3.70\times10^{0}$ & $1.12\times10^{1}$\tabularnewline
\hline 
$500\pm50$ & $8.68\times10^{0}$ & $6.14\times10^{-2}$ & $1.50\times10^{0}$ & $4.96\times10^{0}$\tabularnewline
\hline 
$700\pm70$ & $2.59\times10^{0}$ & $2.72\times10^{-2}$ & $7.25\times10^{-1}$ & $1.97\times10^{0}$\tabularnewline
\hline 
$900\pm90$ & $1.21\times10^{0}$ & $1.24\times10^{-2}$ & $5.13\times10^{-1}$ & $1.18\times10^{0}$\tabularnewline
\hline 
\end{tabular}
\end{table}

The final states result from the decays of $W$ bosons for $W^{+}W^{-}+3b_{jet}$
(where we assume at least one $W$ boson decays leptonically or both
$W$ bosons decay leptonically). In Table \ref{tab:table6}, the statistical
significances $S/\sqrt{B}$ (where $S$ is the number of signal events
and $B$ is the number of background events) for the integrated luminosity
of $L_{int}=500$ fb$^{-1}$, and different types (THDM-I and THDM-II)
and parametrizations (PI and PII) of the model are presented.

\begin{table}
\protect\caption{The statistical significances $S/\sqrt{B}$ at integrated luminosity
$L_{int}=500$ fb$^{-1}$ for different types (THDM-I and THDM-II)
and parametrizations (PI and PII) of the model. The numbers in the
parenthesis show the results for the channel both $W$-boson decay
leptonically. \label{tab:table6}}

\begin{tabular}{|c|c|c|c|c|}
\hline 
Model ($\tan\beta=1$) & \multicolumn{2}{c|}{THDM-I} & \multicolumn{2}{c|}{THDM-II}\tabularnewline
\hline 
$m_{tb}$ (GeV) & PI & PII & PI & PII\tabularnewline
\hline 
\hline 
$300\pm30$ & 402.65 (230.73) & 612.95 (351.24) & 395.02 (226.35) & 553.09 (316.93)\tabularnewline
\hline 
$500\pm50$ & 89.53 (51.30) & 324.52 (185.96) & 95.45 (54.69) & 338.11 (193.75)\tabularnewline
\hline 
$700\pm70$ & 26.56 (15.22) & 164.31 (94.16) & 18.76 (10.75) & 124.56 (71.38)\tabularnewline
\hline 
$900\pm90$ & 6.50 (3.73) & 91.53 (52.45) & 5.73 (3.61) & 69.53 (39.84)\tabularnewline
\hline 
\end{tabular}
\end{table}

\begin{figure}
\includegraphics[scale=0.8]{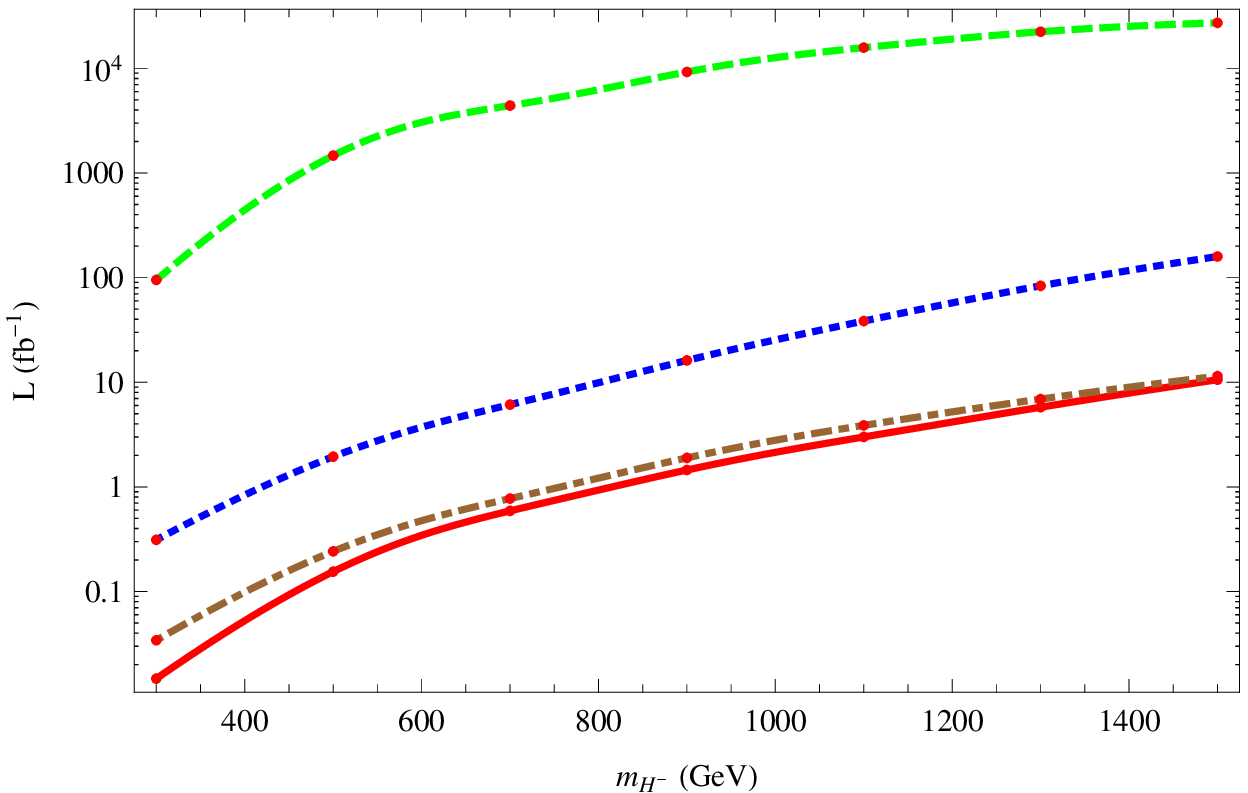}

\protect\caption{The luminosity need to obtain a 3$\sigma$ significance depending
on the mass of charged Higgs boson and different $\tan\beta$ values
(solid line for $\tan\beta=1$, dashed line for $\tan\beta=7$, dotted
line for $\tan\beta=30$, dot-dashed line for $\tan\beta=50$) within
the THDMs. From the decay modes $t\bar{t}b(\bar{b})\to3b_{jet}+2j+l+MET$,
the final state is accounted for at least 3 $b$-jets, 2 light jets,
single charged lepton, and missing transverse momentum. \label{fig:fig25}}
\end{figure}

\begin{figure}
\includegraphics[scale=0.8]{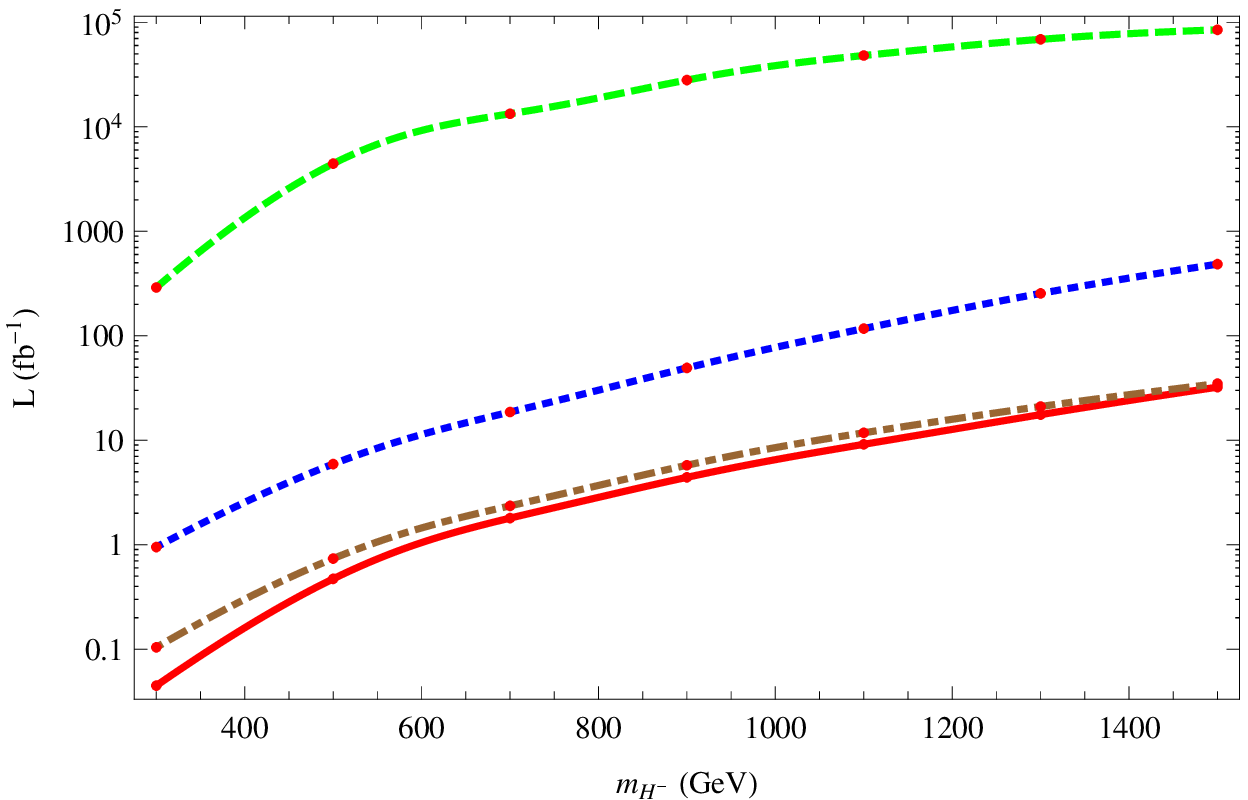}

\protect\caption{The same as Fig. \ref{fig:fig25}, but for the decay modes $t\bar{t}b(\bar{b})\to3b_{jet}+2l+MET$,
the final state is accounted for at least 3 $b$-jets, 2 opposite
charged leptons and missing transverse momentum. \label{fig:fig26}}
\end{figure}

In Figures \ref{fig:fig25} - \ref{fig:fig26}, we present the luminosity
requirement for the signal observability depending on the mass of
charged Higgs boson in the channel - one $W$ boson decays hadronically
while the other decays leptonically (channel - both $W$ bosons decay
leptonically) for single production of charged Higgs boson within
the model framework of THDM-II (parametrization PII) at FCC-hh with
$\sqrt{s}=100$ TeV. The results are shown in different types of lines
for different $\tan\beta$ values.

\section{Conclusions}

Possible extensions of the Higgs sector can be searched for a wide
range of parameter space in the high energy proton-proton collisions.
The ongoing searches at the LHC rely on specific production and decay
mechanism that occupy only a part of the complete model parameter
space. The decay modes of the Higgs bosons can be well similar to
the background reactions from top and bottom quarks and other sources.
If the single production of charged Higgs boson associated with top
quark is observed at the LHC, one of the following questions is to
identify the $H^{-}t\bar{b}$ interaction. The studies on the observables
related to the angular distribution of charged lepton in the final
state and the forward-backward asymmetry can be found in \cite{Gong2014}
and references therein. Even it seems challenging to measure precisely
due to the large hadronic background and systematic uncertainties,
we look forward to its exploitation in precision LHC physics and FCC
physics scenarios. It is shown that with an integrated luminosity
of $500$ fb$^{-1}$ at the center of mass energy $\sqrt{s}=100$
TeV of FCC-hh collider, the signal can be distinguished from the background
for the charged Higgs boson mass up to 1 TeV for a large parameter
space of two Higgs doublet model.

\end{document}